\documentclass[a4paper, 11pt]{article}
\usepackage{float}
\date{March  2009}
\usepackage{epsfig}
\newcommand{\be}{\begin{equation}}
\newcommand{\ee}{\end{equation}}
\newcommand{\ba}{\begin{eqnarray}}
\newcommand{\ea}{\end{eqnarray}}
\newcommand{\bi}{\begin{itemize}}
\newcommand{\ei}{\end{itemize}}

\newcommand{\la}{\label}

\def\a{\alpha}      \def\b{\beta}         \def\G{\Gamma}
\def\d{\delta}        
          \def\l{\lambda}

     \def\s{\sigma}

\newcommand\Tr{\mbox{Tr}\, }

\begin{document}

\begin{titlepage}
%\begin{flushright}
%RBRC 1025\,,\,\,NIKHEF 2013-017
%\end{flushright}

\begin{centering}
\vfill

 \vspace*{2.0cm}
{\bf \Large Matrix Models for Deconfinement and Their Perturbative Corrections}

\vspace{2.0cm}

\centerline{\bf Yun Guo\footnote{yunguo@mailbox.gxnu.edu.cn}}

\vspace{0.5cm}
\centerline{Physics Department, Guangxi Normal University, 541004 Guilin, China}
\centerline{Departamento de F\'isica de Part\'iculas, Universidade de Santiago de Compostela,}
\centerline{E-15782 Santiago de Compostela, Galicia, Spain}

\vspace*{2.0cm}

\end{centering}
\vfill
\centerline{\bf Abstract }
\vspace{0.1cm}
\noindent
Matrix models for the deconfining phase transition in $SU(N)$ gauge theories have been developed in recent years. With a few parameters, these models are able to reproduce the lattice results of the thermodynamic quantities in the semi-quark gluon plasma(QGP) region. They are also used to compute the behavior of the 't Hooft loop and study the exceptional group $G(2)$. In this paper, we review the basic ideas of the construction of these models and propose a new form of the non-ideal corrections in the matrix model. In the semi-QGP region, our new model is in good agreement with the lattice simulations as the previous ones, while in higher temperature region, it reproduces the upward trend of the rescaled trace anomaly as found in lattice which, however, can not be obtained from the previous models. In addition, we discuss the perturbative corrections to the thermal effective potential which could be used to systematically improve the matrix models at high temperatures. In particular, we provide, for the first time, an analytical proof of the relation between the one- and two-loop effective potential: two-loop correction is proportional to the one-loop result, independent of the eigenvalues of the Polyakov loop. This is a very general result, we prove it for all classic groups, including $SU(N)$, $SO(2N+1)$, $SO(2N)$ and $Sp(2N)$.

\vfill

 \end{titlepage}

\setcounter{footnote}{0}

%\begin{document}
\section{INTRODUCTION}\la{intro}

Over the past thirty years, one of the most important aims in high
energy nuclear physics is to explore a new form of matter -- the Quark Gluon Plasma
(QGP), which has been predicted by the lattice Quantum Chromodynamics (QCD)
at finite temperature $T$. It is believed that under the extremely hot and dense condition created in the heavy-ion experiments, hadrons dissolve into a gas of almost free quarks and gluons. The understanding of the transition to the QGP phase at high temperature is expected to progress significantly as a
result of the Relativistic Heavy Ion Collider at Brookhaven and the Large Hadron Collider at CERN.

In this paper, we will focus on the studies of pure $SU(N)$ gauge theories in the deconfined phase. At the asymptotic high temperature, one could calculate the thermodynamics from first principles. However, in the semi-QGP region\cite{semiqgp1,semiqgp2,semiqgp3,semiqgp4}, the computation is very challenging. A useful technique is to perform the Monte-Carlo simulations on the lattice which gives insight into parts of the theory inaccessible by other means.

Lattice results for the thermodynamics of pure $SU(N)$ gauge theories show a rich variety of unexpected behavior near the critical temperature $T_c$. In particular, we are interested in the rescaled conformal anomaly, which is defined as
\be\la{ta}
\tilde \Delta (T) = \frac{e(T)-3 p(T)}{(N_c^2-1)T_c^2 T^2}=\frac{T^3}{(N_c^2-1)T_c^2}\frac{\partial}{\partial T}\bigg(\frac{p(T)}{T^4}\bigg)\,,
\ee
where $e(T)$ and $p(T)$ are the energy density and pressure, respectively. The lattice simulations show that
the rescaled conformal anomaly is roughly a constant in the temperature region from $1.2 T_c$ to $4 T_c$\cite{umeda2009,panero2009,datta2010,Borsanyi:2012ve}. This demonstrates that $\tilde \Delta (T)$ is not dominated by a constant ``bag pressure''.

In order to mimic this behavior, one can add a non-ideal term $\sim T^2$ in the pressure, as did in the previous matrix models\cite{mmo,Dumitru:2010mj,Dumitru:2012fw}. Roughly speaking, the dominate contribution to the rescaled conformal anomaly comes from the term $\sim T^2$ above $1.2T_c$.
%In the narrow region up to $1.2T_c$, there is a rapid decreasing of the background fields, their values are approximately zero above $1.2T_c$.
However, what is the fundamental reason for the appearance of the term $\sim T^2$ in the correction is still an open question. In principle, one can also consider some other possible forms for the non-ideal term in the pressure. In addition, the latest Wuppertal-Budapest lattice data\cite{Borsanyi:2012ve} for pure gauge theory shows that the rescaled conformal anomaly exhibits an upward trend, starting at temperatures above approximately $4T_c$\footnote{The WHOT-QCD results\cite{umeda2009} also have this trend, however, it starts at even lower temperatures.}. Of course, such an behavior can not be described by a term $\sim T^2$ in the pressure. The current matrix models are reliable in the whole semi-QGP region, {\em i.e.}, from $T_c$ to about $4 T_c$.

For very high temperature region, in principle, one can use the perturbative approach to compute the thermodynamics. The next-to-next-to-leading order (NNLO) results from ¡°hard thermal loop¡± (HTL) resummed perturbation theory agree with the lattice data of the rescaled conformal anomaly at temperatures on the order of $8 T_c$\cite{Andersen:2011ug}. One can expect that higher order corrections may further improve the agreement and the perturbation prediction works down to even lower temperatures where the matrix model works. However, the calculations would be rather complicated.

Although the commonly used strategy to describe the behavior of $\tilde \Delta (T)$ in the semi-QGP region is the inclusion of the non-ideal term $\sim T^2$ in the pressure, we will propose another possible solution in this paper. As an alternative, our new proposed model is able to describe the thermodynamics in the semi-QGP region as well as the present matrix models. Furthermore, it also predicts the upward trend starting at temperatures about $4T_c$ as found in the lattice simulation. Our results show that the discrepancy of the rescaled conformal anomaly between the new model and lattice data appears around the temperature of $8 T_c$ where the perturbation theory becomes reliable.

There is a common feature in the matrix models discussed in Refs.\cite{mmo,Dumitru:2010mj,Dumitru:2012fw}. All of them use the one-loop thermal effective
potential as the ideal contributions\cite{Gross:1980br,Weiss1,Weiss2}. As a result, the behavior of the matrix model at high temperatures can be systematically improved by adding the perturbative loop corrections. The effective potential is simply the traditional path integral over the
gauge fields subject to a constraint~\cite{O'Raifeartaigh:1986hi}. This constraint is that the integration is done while preserving the value of the Polyakov loop at some fixed value. Such a procedure generates a probability
distribution for the eigenvalues determined by the fixed value.

The one-loop effective potential has its minima when the value of Polyakov loop equals to $\pm 1$ and the pressure calculated from the minimum of the potential equals the known perturbative pressure calculated at vanishing background field. One important goal to compute the loop corrections is to study how the eigenvalue distribution will be affected.

At present, only the two-loop perturbation corrections have been done\cite{KorthalsAltes:1993ca}. The result shows that the eigenvalue distribution doesn't change and the two-loop effective potential is simply a multiplicative and background independent renormalization of the one-loop result\cite{Dumitru:2013xna},
\be
\la{rela}
\frac{\Gamma^{(2)}}{\Gamma^{(1)}}=-\frac{5g^2C_2(A)}{16\pi^2}\,,
\ee
where $C_2(A)$ is quadratic Casimir invariant in the adjoint representation.

The above relation was first found for the $SU(N)$ groups along straight paths from the origin to the degenerate $Z(N)$ minima\cite{strline1,strline2}. These paths run along the edges of the $SU(N)$ Weyl chamber. However, the minimum of the potential does not exactly follow a straight path as a function of temperature. We should also study if this simple relation still holds inside the Weyl chamber. Aside from $SU(N)$, it is also important to perform the perturbative two-loop calculations for all other classical gauge groups, including the exceptional group $G(2)$\cite{Dumitru:2012fw,Dumitru:2013xna}.

In Ref.\cite{Dumitru:2013xna}, we show that (\ref{rela}) is a very general result for the classic groups including $G(2)$, which holds not only along the edge of the Weyl chamber but also inside. However, by using the MATHEMATICA program, only some specific cases (up to $N=5$) of (\ref{rela}) have been verified. For large N, such a computation becomes rather time-consuming. An analytical proof of the simple relation between one- and two-loop effective potential is still needed. In the second part of this paper, we will give an analytical proof of (\ref{rela}) based on the relations of the Bernoulli polynomials. This proof will finally complete the calculation of the two-loop perturbation corrections to the thermal effect potential which has been studied for about twenty years.

The rest of the paper is organized as the following: in Sec.~\ref{review}, we review the basic idea of the matrix model which was first proposed by Meisinger, Miller and Ogilvie in Ref.\cite{mmo}. We also briefly discuss the improvements of this model. These improvements make the matrix models more reliable for an quantitative comparison to the lattice data. For more details, we refer the readers to Refs.\cite{mmo,Dumitru:2010mj,Dumitru:2012fw}.  In Sec.~\ref{new model}, we propose a new non-ideal correction in the pressure. Although the form looks very different from those in Refs.\cite{Dumitru:2010mj,Dumitru:2012fw}, our new model also predicts the thermodynamics in the semi-QGP region very well. In addition, we also study the upward trend of the rescaled trace anomaly above $4 T_c$ with the new model. In Sec.~\ref{sec:twoloop}, we discuss the two-loop perturbative corrections to the thermal effective potential and provide an analytical proof of the relation given in (\ref{rela}). We prove it for all the classic groups, including $SU(N)$, $SO(2N)$, $SO(2N+1)$ and $Sp(2N)$. The last section contains the conclusions and outlook.

\section{Review of the Matrix Models}\la{review}

The original matrix model was first proposed by Meisinger, Miller and Ogilvie. Considering the 1-loop effective potential in a constant classic background with a dispersion relation $\omega_k=\sqrt{{\bf k}^2+M^2}$ for the gauge bosons, the model is then given by the first two terms in the high temperature expansion of such an effective potential\cite{mmo}
\ba
\la{mmo}
{\cal V}&=& -(N^2-1)\frac{\pi^2 T^4}{45}+\frac{2\pi^2T^4}{3}\sum_{a,b=1}^{N} q_{ab}^2(1-|q_{ab}|)^2+(N^2-1)\frac{T^2M^2}{12}\nonumber \\&-&\frac{T^2M^2}{2}\sum_{a,b=1}^{N} |q_{ab}|(1-|q_{ab}|)\, .
\ea
The classical background field is taken to be diagonal by a gauge rotation and we have
\be
(A_0^{cl})_{ab}=\frac{2\pi T}{g}q_a\delta_{ab}\, .
\ee
$q_{ab}$ is defined as $q_{ab}=q_a-q_b\,, {\rm modulo}\,\, 1$. In this background field the Wilson line is defined as

\be
{\bf L}= {\cal P} {\rm exp} \bigg( i g\int_{0}^{1/T} A_0^{cl} d \tau \bigg)\, ,
\ee
and the Polyakov loop in the fundamental representation is
\be
\ell = \frac{1}{N} {\rm Tr} {{\bf L}}\, .
\ee

In the confined phase, the potential favors $\ell=0$. While at asymptotic high temperature, $\ell=1$. The lattice simulations show that the value of the Polyakov loop changes near the critical temperature $T_c$.  Without the additional mass term in the potential model, the effective potential has its minima when $q_a=0$(or
a $Z(N)$ equivalent state) which indicates $\ell=1$ at all temperatures. In fact, even including the two-loop perturbative corrections, we still have $q_a=0$ as the vacuum. In order to model the transition to deconfinement, as shown in (\ref{mmo}), we need to include the terms containing the mass scale $M$ which can be treated as the non-ideal corrections to the perturbative contributions.

When the effective potential attains at its minimum, we can identify ${\cal V}$ as the free energy which is also equal to the minus pressure. By requiring the effective potential is stationary with respect to the eigenvalues, one can determine the values of $q_a$'s at a given temperature. The only parameter $M$ in the potential model in (\ref{mmo}) is fixed by requiring the phase transition happens at the critical temperature $T_c$.\footnote{In Ref.\cite{mmo}, the parameter $M$ is considered as a temperature independent constant.} As a result, there is no free parameter in this model. Therefore, it is also called the $0-$parameter matrix model.

The predicted thermodynamics can be found in Ref.\cite{mmo}. However, for a quantitative agreement between the model and lattice data, it needs to be improved further. In addition, such a model will have a negative pressure near $T_c$ which is not physical.

In order to get a better fit to the lattice data, in Ref.\cite{Dumitru:2010mj}, an improved matrix model was proposed. The ideal contribution is unchanged as compared to (\ref{mmo}). Using the notations in Ref.\cite{Dumitru:2010mj}, we denote it as ${\cal V}_{pt}$,
\be
\la{vpt}
{\cal V}_{pt}= -(N^2-1)\frac{\pi^2 T^4}{45}+\frac{2\pi^2T^4}{3}\sum_{a,b=1}^{N} q_{ab}^2(1-|q_{ab}|)^2\, .
\ee
The corresponding non-ideal corrections take the following form
\be
\la{vnt}
{\cal V}_{non}= -\frac{4\pi^2}{3} T^2 T_c^2\bigg(\frac{c_1}{5}  V_1({\bf q}) + c_2 V_2({\bf q})- \frac{N^2-1}{60} c_3\bigg)\, ,
\ee
where we define $V_1({\bf q})\equiv \frac{1}{2}\sum_{a,b=1}^{N} |q_{ab}|(1-|q_{ab}|)$ and $V_2({\bf q})\equiv \frac{1}{2}\sum_{a,b=1}^{N} q_{ab}^2(1-|q_{ab}|)^2$. Under the conditions that $c_2=0$ and $c_3=c_1$, the above potential is restored to the one given by (\ref{mmo}) with $c_1=15 M^2/(4 \pi^2 T_c^2)$.
In the improved model, the ratio between $c_1$ and $c_3$ is kept unfixed which is equivalent to adding a term that is independent of the background field and simply proportional to $\sim T^2$. All the parameters $c_1$, $c_2$ and $c_3$ are temperature independent constants. In order to avoid the pressure being negative near the critical temperature, its value at $T_c$ is forced to be zero in the improved model. As a result, there is only one free parameter left which can be determined by fitting the lattice data of the pressure.
%\footnote{However, in Ref., the free parameter is fixed in a different way and we will come to this point later.}.

Furthermore, to get the correct latent heat at the critical temperature, a ``bag" constant term was introduced in the matrix model\cite{Dumitru:2012fw}. Namely, we can replace the parameter $c_3$ in (\ref{vnt}) with $c_3(\infty)+(c_3(1)-c_3(\infty))/t^2$ where the variable $t$ is equal to $T/T_c$ by definition. The improved models\footnote{The improved model proposed in Ref.\cite{Dumitru:2010mj} is also called one-parameter model, while the model with a ``bag" constant in Ref.\cite{Dumitru:2012fw} is called two-parameter model.} are in good agreement with the lattice data and have been used in other studies\cite{robwork1,robwork2,robwork3,robwork4,robwork5,robwork6}.

\section{A New Approach to Improve the Matrix Model}\la{new model}

The original model in (\ref{mmo}) is very simple which contains no free parameter and can be used to describe the thermodynamics for only a qualitative purpose. Therefore, one should consider the improvements for a quantitative description of the lattice data. As we discussed above, the models with a new term $\sim T^2 q_{ab}^2(1-|q_{ab}|)^2$ significantly improve the results. In this section, we propose another different approach to improve the original matrix model which also leads to a perfect fit to the lattice data in the semi-QGP region.

In Refs.\cite{Dumitru:2010mj,Dumitru:2012fw}, the study of the thermodynamics is focused on a temperature region from $T_c$ to $3T_c$. However, as mentioned in the introduction, the rescaled trace anomaly has an upward trend starting at some higher temperature. The previous matrix models fail to reproduce this behavior. With our new model, we will also study the rescaled trace anomaly at higher temperatures.

As before, the ideal contribution in the new model is still given by (\ref{vpt}). However, instead of the inclusion of the term $\sim T^2 q_{ab}^2(1-|q_{ab}|)^2$ in the non-ideal contributions, we consider the temperature dependence of the mass scale $M$ in the original model. Although the above mentioned models treat $M$ as a constant, its temperature dependence was already considered in the quasi-particle models\cite{quasi1,quasi2,quasi3}. Of course, the exact form of the $T$-dependence near the critical temperature can not be calculated from first principles. In this work, we assume the following form for the mass scale\footnote{In some quasi-particle models, the mass $M(T)\sim g(T) T$ matches the perturbative results at very high temperatures. However, our assumption does not satisfy this property. Notice that our considerations focus on the non-perturbative region, in particular, we consider the temperature region from $T_c$ to $8T_c$ in this paper.}
\be\la{mass}
M^2(T) = c_1^{\prime\prime} g^2(T,c_2^\prime) T T_c\, ,
\ee
where $g(T,c_2^\prime)$ is the running coupling which will be considered at the one-loop level in the following,
\be
g^2(T,c_2^\prime) = \frac{24 \pi^2}{11 N\, {\rm ln}(c_2^\prime t)}\, .
\ee
Here, $c_2^\prime$ is another constant parameter that needs to be fixed.
%Notice that we concentrate on the temperature region from $T_c$ to several times $T_c$, therefore, the above temperature dependence of the mass scale is expected to be applicable only in this region.
With this assumption, the non-ideal terms in the original matrix model in (\ref{mmo}) becomes
\be
(N^2-1)\frac{c_1^{\prime\prime} g^2(T,c_2^\prime) T^3 T_c}{12}-c_1^{\prime\prime} g^2(T,c_2^\prime) T^3 T_c V_1({\bf q})\, .
\ee
Instead of $\sim T^2$, the temperature dependence $\sim g^2(T,c_2^\prime)T^3$ appears in the above equation. It turns out to be responsible for the explanation of the upward trend of the rescaled trace anomaly above $4 T_c$. Similar as the $\sim T^2$ terms in the previous matrix models, these non-ideal contributions can be considered as the high temperature corrections to the ideal contributions which are at the order of $\sim T^4$.

As in Ref.\cite{Dumitru:2012fw}, to get the correct behavior at the critical temperature, we should also include two terms that are independent on the background field. Finally, the non-ideal terms in our new model can be expressed as
\be\la{newnpt}
{\cal V}_{non}=-\frac{4\pi^2}{3} T^3 T_c\bigg(\frac{c_1^\prime g^2(T,c_2^\prime)}{5}V_1({\bf q})- \frac{N^2-1}{60}(c_1^\prime g^2(T,c_2^\prime)+\frac{c_3^\prime}{t}+\frac{c_4^\prime}{t^3})\bigg)\, .
\ee
Here, for convenience, we define $c_1^{\prime\prime}\equiv 4 \pi^2 c_1^{\prime}/15$. The four parameters $c_i^\prime$ with $i=1,2,3,4$ are temperature independent constants. By comparing with (\ref{vnt}), we find that (\ref{newnpt}) can be obtained with the following replacements,
\ba\la{rep}
&& c_1\rightarrow c_1^\prime g^2(T,c_2^\prime) t\, ,\quad\quad c_2\rightarrow0\,, \quad\quad c_3\rightarrow c_3(\infty)+(c_3(1)-c_3(\infty))/t^2\, , \\ \nonumber
&&{\rm with}\quad c_3(\infty)\rightarrow c_1^\prime g^2(T,c_2^\prime) t + c_3^\prime\quad{\rm and}\quad c_3(1)-c_3(\infty)\rightarrow c_4^\prime\,.
\ea
The procedure to fix all the parameters follows the same idea as in the previous work\cite{Dumitru:2010mj,Dumitru:2012fw}.
%In fact, with the requirements for the model discussed in Sec. \ref{review}, there is only one free parameter which is fixed by fitting the lattice data of the pressure.
To proceed further, we parameterize the eigenvalues of the Polyakov loop $q_a$'s under the straight line {\em ansatz}:
\be\la{ansatz}
q_a=\frac{N-2a+1}{2N}s\,,
\ee
which assumes the eigenvalues have constant spacing and automatically satisfies the constraint $q_1+q_2+\cdots+q_N=1$ for $SU(N)$ gauge group. In the above equation, $0\le s \le 1$. The confining vacuum corresponds to $s=1$ while the perturbative vacuum is at $s=0$. One can also check that with this parametrization, the value of Polykov loop is real. In fact, we can always consider a real-valued Polyakov loop under a global $Z(N)$ rotation.

Notice that for $N>3$, the exact solution for the $q_a$'s doesn't satisfy this straight line {\em ansatz}. However, the deviation from the straight line is very small\cite{Dumitru:2012fw}. For simplicity, we adopt the {\em ansatz} (\ref{ansatz}) in this section.

Then we can perform the sums in the potential and redefine the potential as
\be\la{newmodel}
{\cal V}_{tot}={\cal V}_{pt}+{\cal V}_{non}=\frac{\pi^2(N^2-1)}{45}T^4 {\cal W}(r,t)\,,
\ee
where
\be\la{w}
{\cal W}(r,t)=\frac{1}{N^2}+\frac{c_3^\prime+c_4^\prime/t^2}{t^2}-\bigg(1+\frac{6}{N^2}-\frac{c_1^\prime g^2(T,c_2^\prime)}{t}\bigg)r^2-2\bigg(1-\frac{4}{N^2}\bigg)r^3
+\bigg(2-\frac{3}{N^2}\bigg)r^4\,.
\ee
Here, $r\equiv 1-s$. The background field $r$ can be obtained by the variational approach
\be
\frac{\partial}{\partial r}{\cal V}_{tot}(r,t)\Big|_{r=r_0(t)}=0\, ,
\ee
and the solution corresponds to the minimum in the deconfined phase is given by
\be\la{r0}
r_0(t)=\frac{1}{8(1-3/(2N^2))}\Big(3(1-\frac{4}{N^2})+\sqrt{25-16(1-\frac{3}{2N^2})\frac{c_1^\prime g^2(T,c_2^\prime)}{t}}\Big)\,.
\ee
In fact, the results in ({\ref{w}) and (\ref{r0}) can be directly read off from Ref.\cite{Dumitru:2012fw} by using the replacements given in (\ref{rep}).

Impose the condition ${\cal W}(0,1)={\cal W}(r_0(1),1)=0$ at the critical temperature, one can get the following relations among these parameters
\be
c_1^\prime g^2(T_c,c_2^\prime)=\frac{(N^2+1)(3N^2-2)}{N^2(2N^2-3)} \, ,\,\,\,c_3^\prime+c_4^\prime=-\frac{1}{N^2}\,.
\ee
An additional constraint on these parameters can be obtained by considering the latent heat which is the jump in the energy density at $T_c$. By construction, both the pressure and the energy density vanish at the confined phase, as a result, the latent heat equals to the rescaled energy density $e(T_c)/T_c^4$ and so the trace anomaly $(e(T_c)-3p(T_c))/T_c^4$.

Based on the matrix model, the straightforward calculation of the latent heat leads to another constraint on the parameters $c_2^\prime$ and $c_4^\prime$. It reads
\be
c_4^\prime=\frac{45{\cal L}(N)}{2\pi^2}-\frac{(N^2-4)^2(3N^4+N^2-2)}{2N^2(2N^2-3)^3 {\rm ln}(c_2^\prime)}-\frac{3N^8-39N^6+110N^4-76N^2+22}{2N^2(2N^2-3)^3}\,,
\ee
where ${\cal L}(N)$ is the latent heat renormalized by the number of perturbative gluons $N^2-1$.
According the the lattice result\cite{datta2010}, we have ${\cal L}(3)=0.209$, ${\cal L}(4)=0.287$ and ${\cal L}(6)=0.342$.

Now there is only one free parameter left which will be fixed by fitting the lattice data of the pressure\cite{datta2010}. In order to make an direct comparison with the previous model, we also consider the $SU(N)$ gauge theories for $N=3,4$ and $6$. The values of the four parameters and the corresponding ``bag" constants $B=\pi^2(N^2-1)c_4^\prime T_c^4/45 $ are summarized in table \ref{para}. \footnote{We choose $T_c=270$ MeV.}
\begin{table}
\begin{tabular}{|c c c c c c|}
\hline
$N$ \quad \quad&\quad  $c_1^\prime$ \quad \quad&\quad \quad $c_2^\prime$ \quad  \quad&\quad\quad  $c_3^\prime$  \quad \quad&\quad \quad $c_4^\prime$ \quad\quad& $B^{1/4}({\rm MeV})$ \\
\hline
3   \quad\quad&\quad    0.1352 \quad \quad&\quad \quad   1.6885    \quad \quad&\quad \quad   -0.3994 \quad \quad&\quad \quad   0.2883  \quad \quad& 228 \\
4   \quad \quad&\quad    0.1989  \quad \quad&\quad \quad   1.8878   \quad \quad&\quad \quad   -0.4079    \quad \quad&\quad \quad   0.3454  \quad \quad&  279 \\
6   \quad \quad&\quad    0.3537  \quad \quad&\quad \quad    2.2342   \quad \quad&\quad \quad   -0.4542    \quad \quad&\quad \quad   0.4264  \quad \quad& 363\\
\hline
\end{tabular}
\caption{The values of the four parameters in the matrix model and the corresponding ``bag" constants for $SU(N)$ gauge theories for $N$=3,4 and 6.}
\label{para}
\end{table}
The ``bag" constant $B$ increases with $N$ and the numerical values have a good agreement with those from the matrix model in Ref.\cite{Dumitru:2012fw}.

\begin{figure}[htbp]
\includegraphics[width=0.5\textwidth]{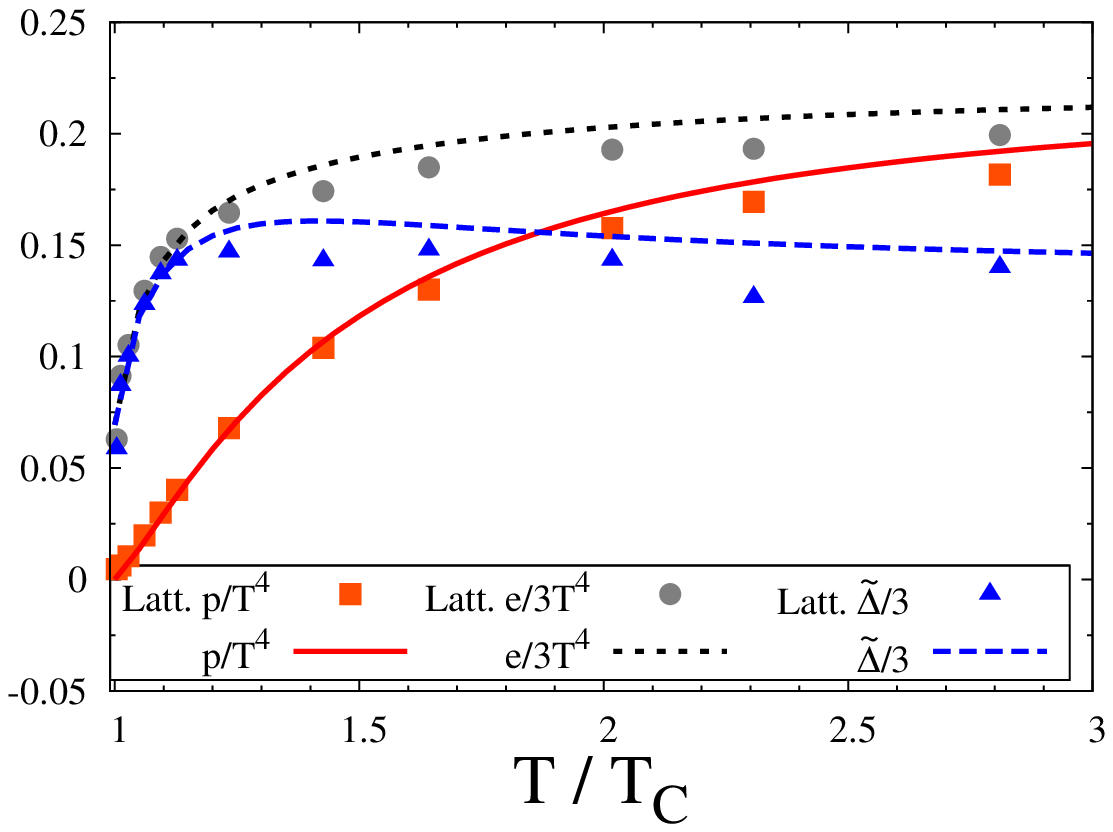}
\includegraphics[width=0.5\textwidth]{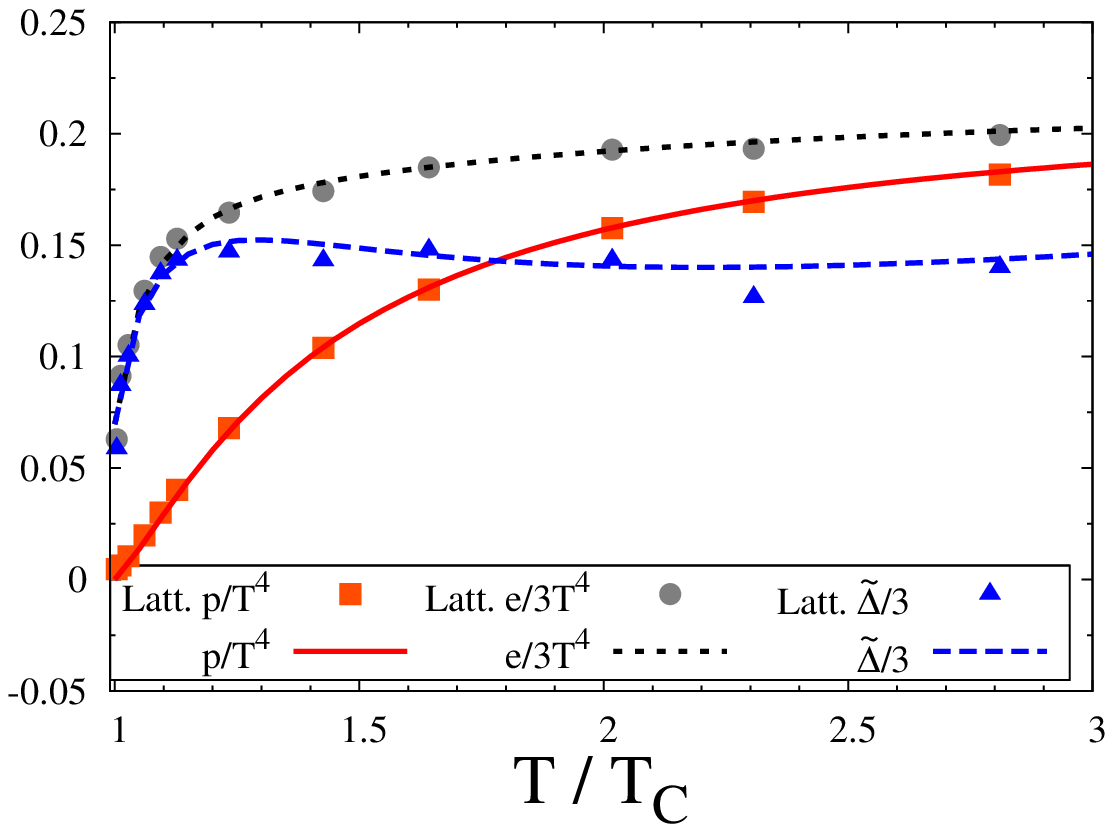}
\vspace*{-.5cm}
\caption{\label{Nc3}
Comparison of the $SU(3)$ thermodynamics obtained from the lattice simulation and matrix models. We consider the dimensionless pressure $p/T^4$, energy density $e/(3T^4)$, as well as one third the rescaled trace anomaly defined in (\ref{ta}). All quantities are also scaled by 1/8. Left: the matrix model is taken from Ref.\cite{Dumitru:2012fw}. Right: the matrix model is taken from (\ref{newmodel}).}
\end{figure}

\begin{figure}[htbp]
\includegraphics[width=0.5\textwidth]{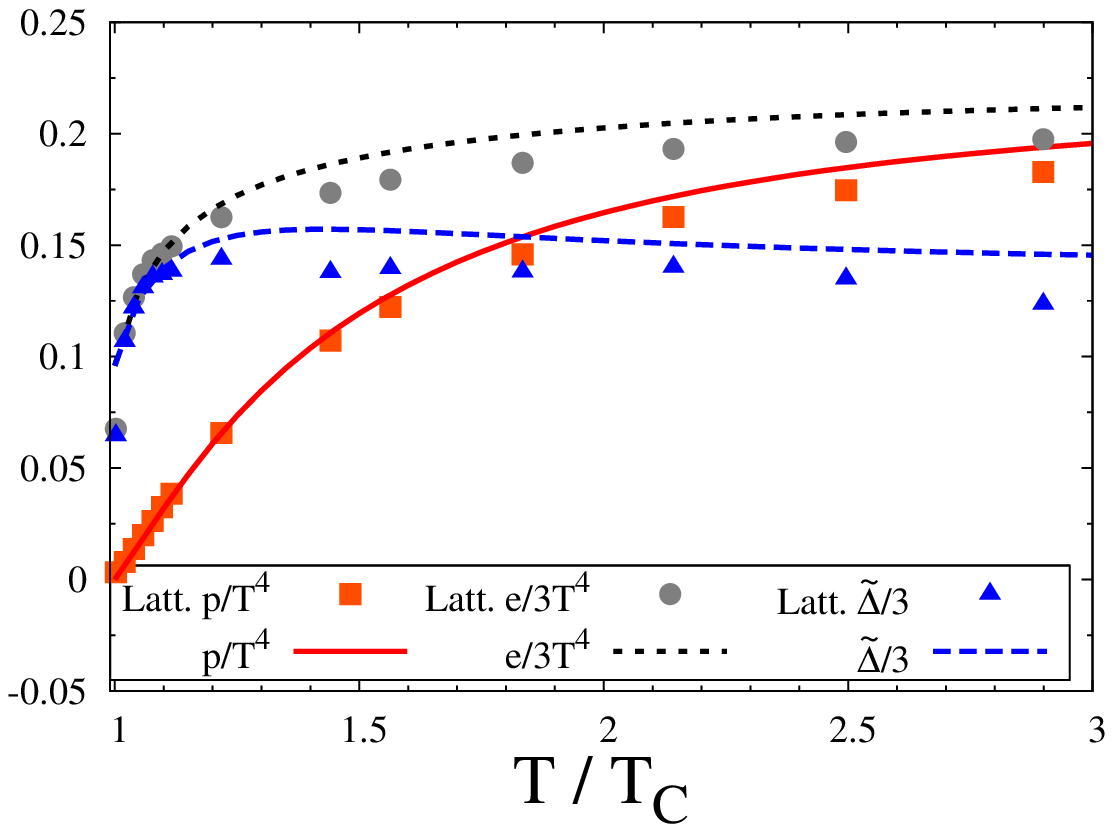}
\includegraphics[width=0.5\textwidth]{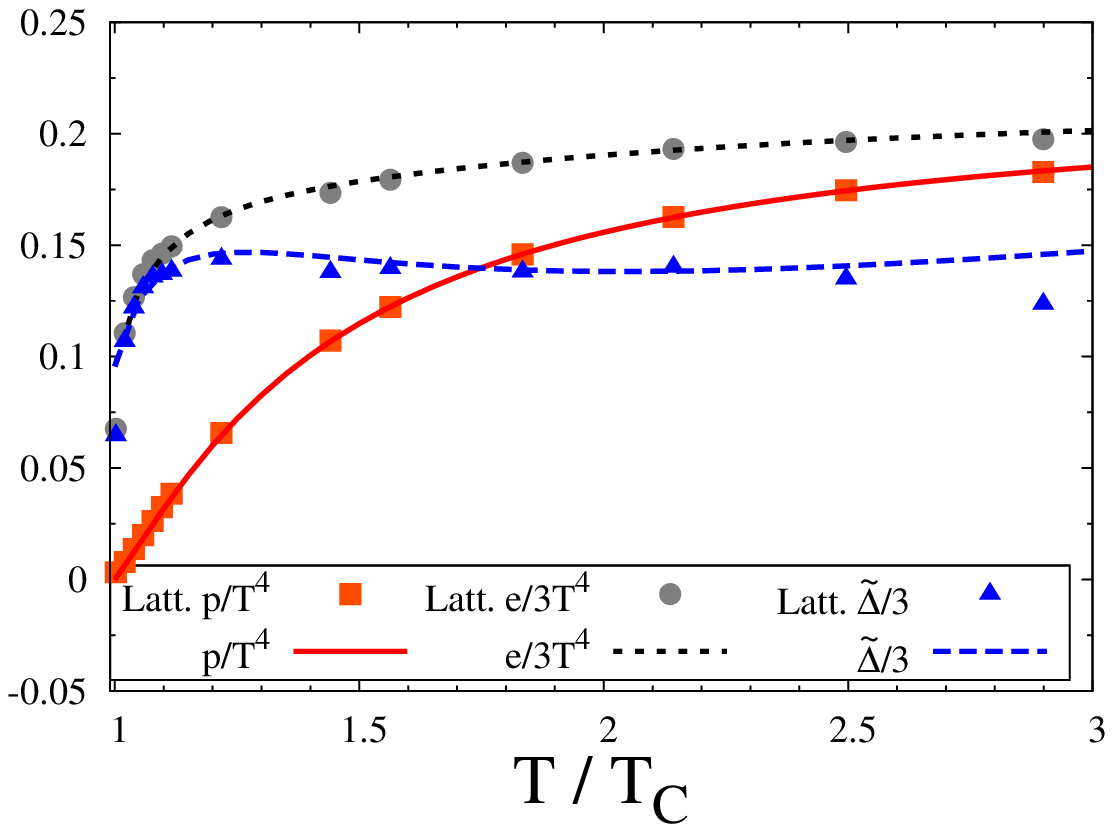}
\vspace*{-.5cm}
\caption{\label{Nc4}
Comparison of the $SU(4)$ thermodynamics obtained from the lattice simulation and matrix models. We consider the dimensionless pressure $p/T^4$, energy density $e/(3T^4)$, as well as one third the rescaled trace anomaly defined in (\ref{ta}). All quantities are also scaled by 1/15. Left: the matrix model is taken from Ref.\cite{Dumitru:2012fw}. Right: the matrix model is taken from (\ref{newmodel}).}
\end{figure}

\begin{figure}[htbp]
\includegraphics[width=0.5\textwidth]{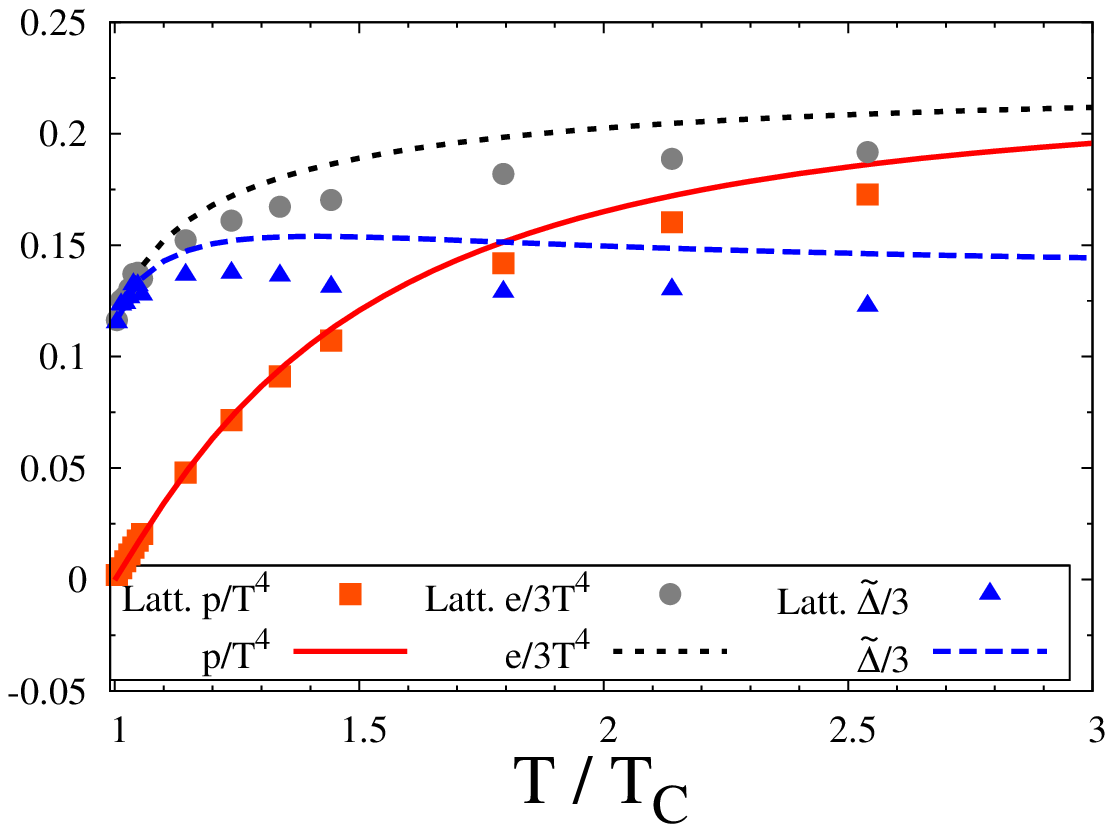}
\includegraphics[width=0.5\textwidth]{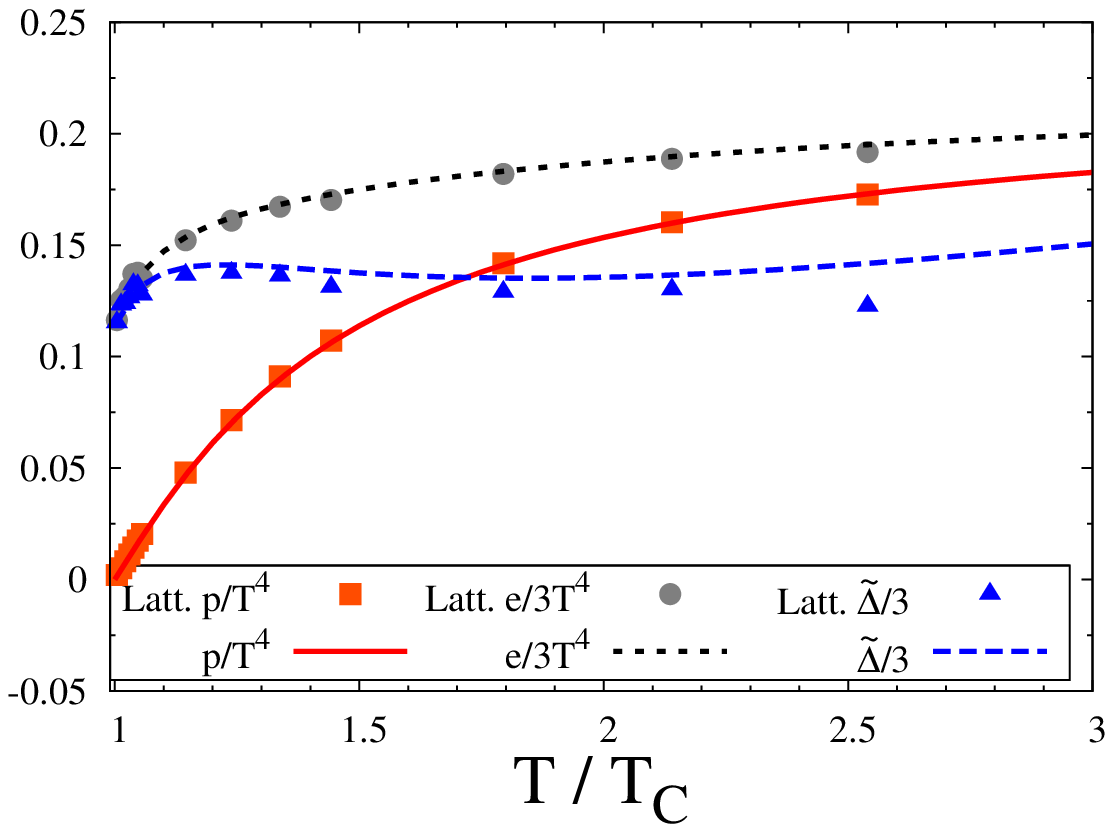}
\vspace*{-.5cm}
\caption{\label{Nc6}
Comparison of the $SU(6)$ thermodynamics obtained from the lattice simulation and matrix models. We consider the dimensionless pressure $p/T^4$, energy density $e/(3T^4)$, as well as one third the rescaled trace anomaly defined in (\ref{ta}). All quantities are also scaled by 1/35. Left: the matrix model is taken from Ref.\cite{Dumitru:2012fw}. Right: the matrix model is taken from (\ref{newmodel}).}
\end{figure}

The comparisons of the $SU(N)$ thermodynamics obtained from lattice simulation and matrix models for $N=3,4$ and $6$ are shown in Figs.(\ref{Nc3}), (\ref{Nc4}) and (\ref{Nc6}), respectively. Thermodynamic quantities considered here are the dimensionless pressure $p/T^4$, energy density $e/(3T^4)$, as well as one third the rescaled trace anomaly defined in (\ref{ta}). Two kinds of the matrix models are included. The left parts of these figures show the results predicted by the matrix model in Ref.\cite{Dumitru:2012fw}, while the right parts show the results from the model given in (\ref{newmodel}).

Although having very different forms, both matrix models appear to reproduce the lattice data reasonably well in the semi-QGP region. In the new model, the complication of the inclusion of a $T$-dependent mass scale is compensated by the absence of the term $\sim T^2 V_2({\bf q})$.

Taking a close look into the figures, we find that for the two-parameter matrix model, deviations from the lattice data become visible at higher temperatures, while for the new model, the agreement is rather good in the entire semi-QGP region. Despite the structures of the non-ideal terms in these models, such a difference is probably induced by the way how one fixes the free parameter. Instead of fitting the lattice data of the pressure as we did in this paper, in Ref.\cite{Dumitru:2012fw}, the authors fixed this parameter by considering the behavior of the rescaled trace anomaly.

As already mentioned in the introdudction, $\tilde\Delta(T)$ is almost constant in the temperature region from $1.2T_c$ to $4T_c$ where the background fields are approximately zero\footnote{As a result, one can take variable $r\approx 1$ in the potential.}, therefore, the contribution to $\tilde\Delta(T)$ is dominated by the term $\sim T^2$ in the potential\footnote{In the two-parameter model, the ``bag" constant term leads to a $\sim 1/T^2$ contribution to $\tilde\Delta(T)$ which is small when $T$ is relatively large.}. Using the value of $\tilde\Delta(T)$ from the lattice simulation, the parameter related to the term $\sim T^2$, which is denoted as $c_3(\infty)$ in Ref.\cite{Dumitru:2012fw}, can be determined. This determination is universal for any $N$ since the numerical result of $\tilde\Delta(T)$ is basically independent on $N$ as indicated by the lattice.

However, the above analysis can not be used in the model considered in this paper. As in the previous models, we also find a rapid fall-off of the background fields near $T_c$ and their values almost vanish above $1.2 T_c$. However, beside the $\sim T^2$ term, the term $\sim T^3 g^2(T,c_2^\prime)$ also contributes to the rescaled trace anomaly. The same behavior of $\tilde\Delta(T)$ is also found in the new model because the contribution to $\tilde\Delta(T)$ coming from the term $\sim T^3 g^2(T,c_2^\prime)$ is also approximately a constant from $1.2T_c$ to $4T_c$. However, as the temperature increases, such a contribution starts to increase with $T$. The same trend is found in the lattice simulation\cite{Borsanyi:2012ve}. On the other hand, with previous matrix models, the change of $\tilde\Delta(T)$ is negligible above $1.2T_c$.

In Fig.(\ref{Nc3large}), we show the thermodynamic quantities for $SU(3)$ up to $8T_c$. The corresponding lattice data can be found in Ref.\cite{Borsanyi:2012ve}. With our new model, the upward trend of the resacled trace anomaly above $4 T_c$ is observed which, however, disappears when using the previous matrix models. Quantitatively, a slight discrepancy exists because the values of the parameters are taken from table \ref{para} and we don't attempt to refix them. The lattice data of the resacled trace anomaly is available for even higher temperatures where, however, the model description fails. As compared to the lattice simulation, a slower increasing with temperature is found from our model. We comment that the temperature dependence of the mass scale in (\ref{mass}) is proposed only at the phenomenological level and our consideration is concentrated in the non-perturbative region. Above $8T_c$, the HTL perturbation theory works very well for the $SU(N)$ gauge theory.

\begin{figure}
\begin{center}
\includegraphics[width=0.6\textwidth]{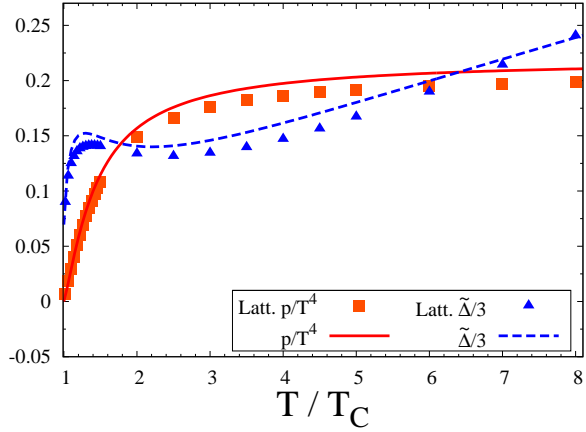}
\end{center}
\vspace*{-.5cm}
\caption{\label{Nc3large}
Comparison of the $SU(3)$ thermodynamics obtained from the lattice simulation and the new matrix model in (\ref{newmodel}). We consider the dimensionless pressure $p/T^4$%, energy density $e/(3T^4)$, as well as
and one third the rescaled trace anomaly defined in Eq.(\ref{ta}) up to $8T_c$. All quantities are also scaled by 1/8. %Left: the matrix model is taken from Ref.. Right: the matrix model is taken from Eq.(\ref{newmodel}).
}
\end{figure}

It is also interesting to see the temperature dependence of the gluon mass $M(T)$ as given in (\ref{mass}). If we parameterize the mass $M(T)$ as $M^2(T)=g^2_{eff} T^2 N/6$ according to the leading order perturbative result, the behavior of  the effective coupling $g^2_{eff}=8\pi^2 c_1^\prime g^2(T, c_2^\prime)/(5 N t)$ is given in Fig.(\ref{geff}). On the right hand side of this figure, we plot the ratio of $M^2(T)$ to $T^2$ which is the effective coupling scaled by a factor $N/6$. It turns out that the effective coupling increases as the temperature approaches to $T_c$ and the ratio of $M^2(T)$ to $T^2$ is essentially independent on $N$. Some visible $N$-dependence only exists in a narrow region close to the critical temperature. Notice that in the two-parameter model, the mass is a constant and the values of $M^2/T_c^2$ are $2.19$, $2.70$, $3.17$ for $N=3,4,6$, respectively.

\begin{figure}[htbp]
\includegraphics[width=0.5\textwidth]{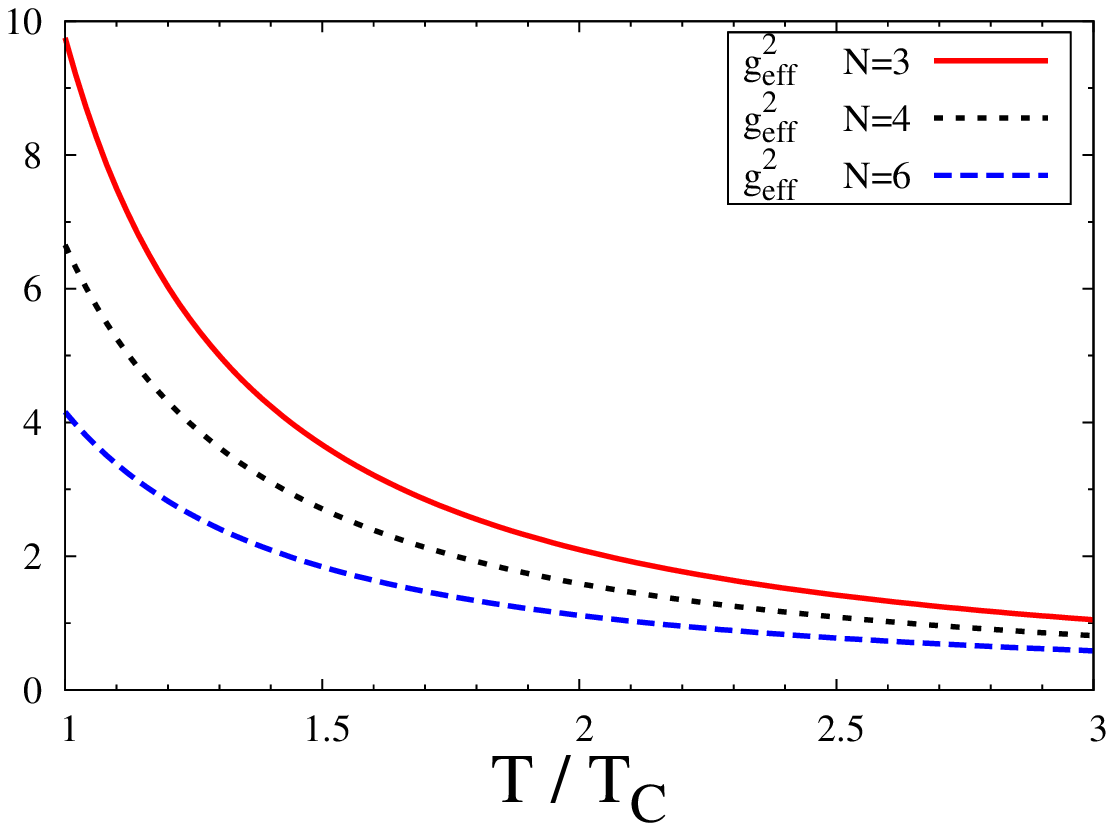}
\includegraphics[width=0.5\textwidth]{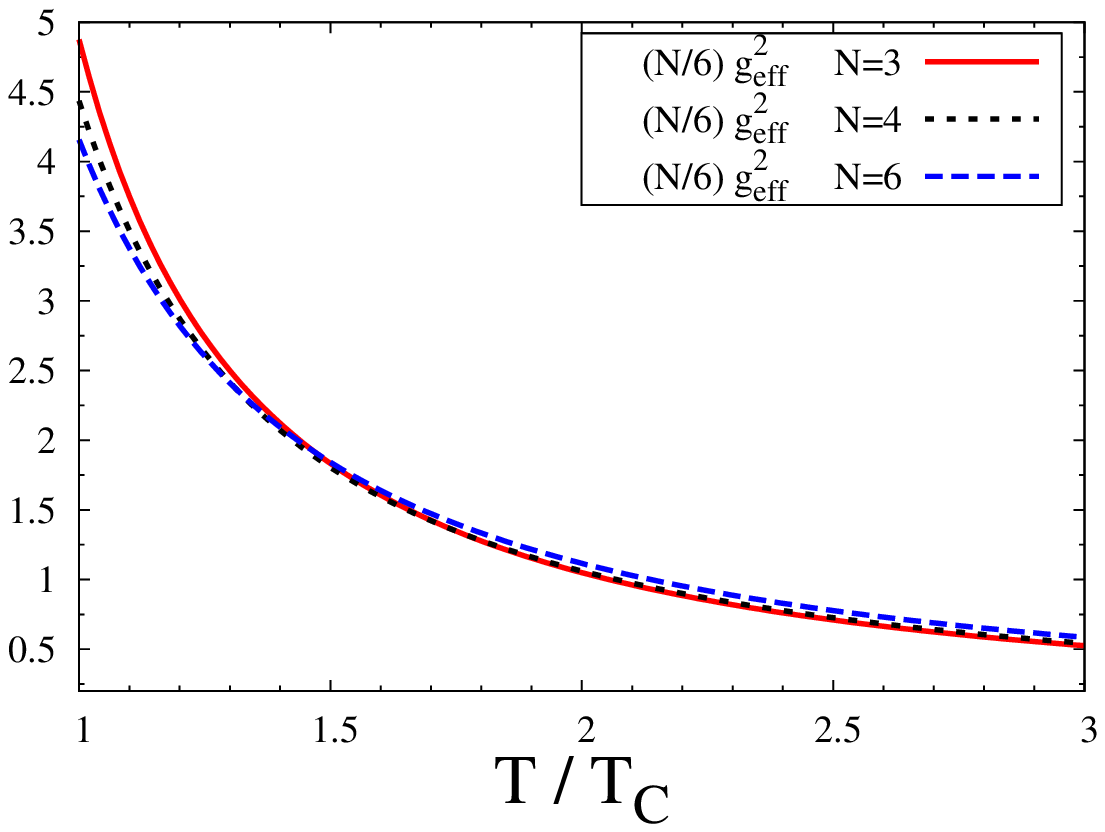}
\vspace*{-.5cm}
\caption{\label{geff}
The temperature dependence of the effective coupling for $N=3,4,$ and $6$. Left: $g^2_{eff}=6 M^2(T)/(T^2 N)$. Right: The ratio of $M^2(T)$ to $T^2$ which equals to $(N/6) g^2_{eff}$.}
\end{figure}

Lastly, we mention that due to the rapid fall-off of the background fields near the critical temperature, the Polyakov loop predicted by the two-paramter matrix model differs from $1$ only in a very narrow region, from $T_c$ to 1.2$T_c$.  However, the renormalized Polyakov loop from the lattice varies over the entire semi-QGP. Although the increasing rate of the loop near $T_c$ found in the new model is relatively smaller as compared to the two-paramter matrix model, the agreement with the lattice results is still not satisfactory.

\section{Two-loop Perturbation Corrections to the Effective Potential}\la{sec:twoloop}

In this section, we discuss the perturbation corrections to the effective potential. %As all the matrix models use the one-loop effective potential ${\cal V}_{pt}$ in (\ref{vpt}) as the ideal contributions, they can be systematically improved by including the higher loop perturbation corrections. Therefore,
The basic properties of these loop corrections are of particular interest as we already mentioned in the introduction. Here, we will focus on the studies of the simple relation between one- and two-loop effective potential as given in (\ref{rela}). Notice that the explicit result of the effective potential up to two-loop is known since long, however, there is nothing in the way one performs the computation that suggests such a simple relation. Previous work can be found in \cite{KorthalsAltes:1993ca,Dumitru:2013xna,strline1,strline2}, however, an analytical proof is not available so far.

Using the explicit form of the effective potential, we will give an analytical proof of (\ref{rela}) in the following. For completeness, we will prove it for all the classic groups, including $SU(N)$, $SO(2N)$, $SO(2N+1)$ and $Sp(2N)$.

We rewrite the effective potential up to two-loop order as
\ba
\la{1loop}
\G^{(1)}&=& -\frac{\pi^2 T^4 d(A)}{45}+\sum_{a}\widehat B_4(q_a)\, ,\\
\la{newuf}
\G^{(2)}_f&=& g^2\sum_{a,b,c}|{f^{a,b,c}}|^2\widehat B_2(q_b)\widehat B_2(q_c)\, ,\\
\la{newoneloopinsertfinal1}
\G^{(2)}_i&=&2 g^2\sum_{d,b,c}f^{d,b,-b}f^{d,c,-c}\widehat B_1(q_{b})\widehat B_3(q_{c})\,.
\ea
In the above equations,  $d(A)$ is the dimension of the
adjoint representation of the group and equals $N^2-1$ for $SU(N)$, $2N^2-N$ for $SO(2N)$, $14$ for $G(2)$,
$2N^2+N$ for both $Sp(N)$ and $SO(2N+1)$.
$f^{a,b,c}$ is the structure constant of the group which can be determined from the commutation relations of the group generators.
The definition of the Bernoulli Polynomials $\widehat B_i(x)$ for $i=1,2,3,4$ is given in Appendix \ref{appa}.
We also divide the two-loop effective potential into two parts. $\G^{(2)}=\G^{(2)}_f+\G^{(2)}_i$.
$\G^{(2)}_f$ denotes the contributions from usual two-loop free-energy diagrams, while $\G^{(2)}_i$ corresponds to the contribution from the insertion diagram\cite{KorthalsAltes:1993ca,Dumitru:2013xna}.
In $\G^{(2)}_f$, the
indices $a$,$b$ and $c$ run over both diagonal and off-diagonal
indices\footnote{The (off-)diagonal indices are related to the indices of the (off-)diagonal generators of the group in Carten space. For more details, we refer the reader to Ref.\cite{Dumitru:2013xna} and the references therein.}. In $\G^{(2)}_i$, each structure constant contains the
diagonal indices $d$, while $b$ and $c$ denote off-diagonal
indices. In addition, we have $E^{-a}\equiv (E^{a})^{\dagger}$ for the off-diagonal generators which defines the ``minus" indices.

In (\ref{rela}), the Casimir invariant is related to the rank of the group $d(r)$ through
\be
C_2(A)\; d(r) = f^{a,b,d}f^{a,b,d}~.
\la{casimirroots}
\ee
For each classic group, we have that $C_2(A)= N-1$ for $SO(2N)$, $N-{1\over 2}$
for $SO(2N+1)$, $N+1$ for $Sp(2N)$, and $C_2(A)=N$ for the $SU(N)$ groups.
%For the $SU(N)$ groups the result~(\ref{twoloopresult}) was in fact known since long for straight
% paths\footnote{However, in general the minimum of the potential does not exactly follow a straight path as a function of temperature~\cite{Dumitru:2012fw}.} from the origin, ${\bf q}=0$, to
%the degenerate $Z(N)$ minima~\cite{Bhattacharya:1992qb}. These paths
%run along the edges of the $SU(N)$ Weyl chamber and a combinatorial
%proof of Eq.~(\ref{twoloopresult}) exists~\cite{Giovannangeli:2002uv}.
%We do not (yet) know how the combinatorics works out inside the Weyl
%chamber.

In order to continue our discussion, some generalities on the classical Lie algebras should be mentioned\footnote{They can be found in Refs.\cite{Dumitru:2013xna,georgi}. Here, for completeness, we will repeat some important aspects.}. For any semi-simple Lie algebra, the commutation relations in the Cartan basis are given by
\ba
~[\vec H,E_{\a}]&=&\vec\a ~E_{\a} \\ \nonumber
~[E_{\a},E_{-\a}]&=&\vec\a \cdot \vec H\\ \nonumber
~[E_{\a},E_{\b}]&=&f^{\a,\b,-\a-\b}E_{\a+\b}\, ,
~\mbox{if}~ \a+\b~\mbox{is a root; if not, it vanishes.}
 \la{eq:appcommrel}
\ea
The diagonal generators in the
fundamental representation are the components of $\vec H$ which are the orthonormal matrices spanning the
Cartan subalgebra. The off-diagonal generators are the orthonormal $E_\a$, labelled by the roots $\a$. They are vectors
in Cartan space.

A root labeled by an off-diagonal index $a$ has $d$ components which are the structure constants $f^{d,a,-a}$. They can be determined by the commutation relation
\be
~[H^d,E_a]=f^{d,a,-a}E_a.
\ee
There is another
kind of structure constants $f^{\a,\b,-\a-\b}$ which connect
off-diagonal generators. These generators are normalized as
\be
\Tr (E_\a E_{-\a})=\Tr (H_d^2)=1/2~.
\ee

The proof of (\ref{rela}) can be achieved by the following two steps. Firstly, by using the corresponding commutation relations of the group generators, we are able to get some simplified expressions for the two-loop effective potential which contain only the Bernoulli polynomials while the structure constants appear in (\ref{newuf}) and (\ref{newoneloopinsertfinal1}) have been get rid of. Secondly, based on a set of relations of the Bernoulli polynomials, the simplified two-loop results are expressed in terms of $\widehat B_4(x)$. Then a direct comparison between the one- and two-loop effective potential proves the renormalization relation given in (\ref{rela}).

In fact, the insertion diagram involves sums over diagonal indices $d$ which
can be performed quite easily as they correspond to inner products
between the corresponding roots. This has already been done in \cite{Dumitru:2013xna}. For later use, the simplified results for
$\Gamma^{(2)}_i$ for each group are list below.

For $SU(N)$, we have
\be
\la{g2sunsim}
\G_i^{(2)}(SU(N))=4 g^2\sum_{ijl}\widehat B_1(q_i-q_j)\widehat B_3(q_i-q_l)\,.
\ee

For the orthogonal groups, the result is the following

\be
\la{g2so2nsim}
\G_i^{(2)}(SO(2N))=2 g^2\sum_{ijl}\bigg(\widehat B_1(q_i+q_j)+
\widehat B_1(q_i-q_j)\bigg) \bigg(\widehat B_3(q_i+q_l)+ \widehat
B_3(q_i-q_l)\bigg)\,,
\ee
and
\ba
\la{g2so2n+1sim}
&\G_i^{(2)}(SO(2N+1))=\G_i^{(2)}(SO(2N))+2
g^2\sum_{i,j}\bigg[(\widehat B_1(q_i+q_j)+\widehat
  B_1(q_i-q_j))\nonumber \\ &\widehat B_3(q_i)+(\widehat
  B_3(q_i+q_j)+\widehat B_3(q_i-q_j))\widehat B_1(q_i)\bigg]
+2 g^2\sum_{i}\widehat B_1(q_i)\widehat B_3(q_i)\,.
\ea

Finally, for $Sp(2N)$,the simplified insertion diagram reads
\ba
\la{g2sp2nsim}
\G_i^{(2)}(Sp(2N))&=&2 g^2\sum_{i}\bigg(\sum_{j}(\widehat
B_1(q_i+q_j)+\widehat B_1(q_i-q_j))+2 \widehat B_1(2
q_i)\bigg)\nonumber \\ &\times& \bigg(\sum_{l}(\widehat
B_3(q_i+q_l)+\widehat B_3(q_i-q_l))+2 \widehat B_3(2 q_i)\bigg)\, .
\ea

Notice that in these results, the constraints on the indices $i\neq j$ and $i\neq l$ apply.

\subsection{Proof for $SU(N)$}
\la{sun}

Starting from (\ref{newuf}) and~(\ref{newoneloopinsertfinal1}),
we are able to calculate the two-loop perturbative correction to the
effective potential. First of all, we need to know the structure
constants. They can be obtained from the commutation relations of the
generators in Cartan basis.

For $SU(N)$, there are $N(N-1)$ off-diagonal generators
$E^{ij}\equiv\l^{ij}$ with $i,j=1,\cdots,N$ and $i \neq j$. The
explicit forms are given by $(\l^{ij})_{kl}=\delta_{ki}\delta_{jl}/\sqrt{2}$. In addition, we
have $N-1$ traceless diagonal generators $H^d\equiv\l^{d}$ with
$d=1,\cdots,N-1$,
\be
\l^{d}=\frac{1}{\sqrt{2d(d+1)}}\,
{\rm diag}(1,1,\cdots\, ,-d,0,0,\cdots\,,0)\, .
\ee
The commutators between diagonal generators are obviously zero. The
nonvanishing commutators we need are\footnote{For $SU(N)$, if a
  typical off-diagonal index $b$ is denoted by $b=ij$, then we have
  $-b=ji$.}
\ba
\la{com1}
[H^{d},E^{ij}]&\equiv&f^{d,ij,lk}E^{kl}=(\l^d_{ii}-\l^d_{jj})
E^{ij}\,,\\{}
[E^{ij},E^{kl}]&\equiv&f^{ij,kl,ts}E^{st}= \frac{1}{\sqrt{2}}(\d_{jk}E^{il}-\d_{il}E^{kj})\,.
\la{com}
\ea
Here, $\l^d_{ii}$ is the $i$th diagonal component of $\l^d$. From
(\ref{com1}) we find that the roots ${\vec \a}^{\,ij}=({\vec
  \l}_{ii}-{\vec \l}_{jj})$. As mentioned before, the $d$th
component of ${\vec \a}^{\,ij}$ is the structure constant
$f^{d,ij,ji}$. In addition, (\ref{com}) indicates that the non-vanishing structure constant is
\be
\la{sunsc}
f^{ij,jk,ki}=\frac{1}{\sqrt{2}}\, .
\ee

%We can define a diagonal matrix,
%\be
%\la{dl}
%\L^{ij}\equiv \frac{{\vec \l}\cdot{\vec \a}^{\,ij}}{({\vec \a}^{\,ij})^2}\, ,
%\ee
%and it is easily to find the following commutator:
%\ba
%[\L^{ij},E^{ij}]&=& E^{ij}\, .
%\ea
%
%Using the explicit form of $E^{ij}$, we get $\L^{ij}=\frac{1}{2} {\rm
%  diag}(0,0,\cdots,1,0,\cdots,0,-1,0,\cdots,0)$, i.e.\ the $i$th
%component is $1$, the $j$th component is $-1$ and all others are zero.

%Taking the square of Eq.~(\ref{dl}) and then the trace on both
%sides, the roots satisfy
%
%\be
%\la{roots}
%({\vec \a}^{\,ij})^2=1\, .
%\ee
%
%In other words, the roots can be written in terms of an orthonormal
%basis $\{{\vec e}_i\}$ spanning an $N$-dimensional space,
%\be
%{\vec \a}^{\,ij}=\frac{1}{\sqrt{2}}({\vec e}_i-{\vec e}_j)\, .
%\ee
%
%Using Eqs.~(\ref{casimirroots}) and~(\ref{roots}), we have $C_2(A)=N$
%for $SU(N)$. Notice that there are $N^2-N$ off-diagonal indices and
%that the rank of $SU(N)$ is $N-1$.
For $SU(N)$, the arguments in the Bernoulli functions have two different cases: for a diagonal index $a$, $q_a = 0$; for an off-diagonal index $a=ij$, $q_a = q_i-q_j$ .

Firstly, we consider (\ref{newuf}) in the case where only one of the indices $a$, $b$ and $c$ is the diagonal index\footnote{If more than one index is diagonal, such a term has no contribution.}. By using the property of the roots ${\vec \a}^{\,ij}=\frac{1}{\sqrt{2}}({\vec e}_i-{\vec e}_j)$ with an orthonormal
basis $\{{\vec e}_i\}$ spanning an $N$-dimensional space, we can get rid of the structure constants and write this kind of contribution as
\be
\la{sungfpart1}
\Gamma_{f}^{(2)}|_{\rm{I}} = g^2\sum_{i\neq j} \bigg(2 \widehat B_2(0) \widehat B_2(q_i - q_j) +
(\widehat B_2(q_i - q_j))^2\bigg)\, .
\ee

The other contribution from (\ref{newuf}) then becomes
\be
\la{uf2}
\Gamma_{f}^{(2)}|_{\rm{II}}=g^2\sum_{mn,ij,kl}|{f^{mn,ij,kl}}|^2\widehat B_2(q_i - q_j)\widehat B_2(q_k - q_l) \, ,
\ee
which corresponds to the case where all the indices $a$, $b$ and $c$ are the off-diagonal indices.
In the above euqations, we symbolically write $\Gamma_{f}^{(2)} \equiv \Gamma_{f}^{(2)}|_{\rm{I}}+\Gamma_{f}^{(2)}|_{\rm{II}}$, the same applies to $\Gamma_{i}^{(2)}$.

Using (\ref{sunsc}), (\ref{uf2}) can be simplified as
\ba
\la{sungfpart2}
\Gamma_{f}^{(2)}|_{\rm{II}}&=& g^2\sum_{mnijl}\widehat B_2(q_i - q_j)\bigg(|{f^{mn,ij,jl}}|^2\widehat B_2(q_j - q_l)+|{f^{mn,ij,nl}}|^2\widehat B_2(q_n - q_l)\bigg)\,\nonumber\\
&=& \frac{g^2}{2}\bigg(\sum_{i\neq j\neq l}\widehat B_2(q_i - q_j)\widehat B_2(q_j - q_l)+\sum_{m\neq n\neq l}\widehat B_2(q_l - q_m)\widehat B_2(q_n - q_l)\bigg)\,\nonumber\\
&=&g^2 \sum_{i\neq j\neq k} \widehat B_2(q_i - q_j)\widehat B_2(q_i - q_k)\, .
\ea
Notice that $i\neq j\neq k$ means $i\neq j$, $i \neq k$ and $j \neq k$.

For later use, we split $\G_i^{(2)}$ given in (\ref{g2sunsim}) into two parts as
\ba
\la{sungipart1}
\G_i^{(2)}|_{\rm{I}}&=& 4 g^2\sum_{i\neq j}\widehat B_1(q_i-q_j)\widehat B_3(q_i-q_j)\,,\\
\la{sungipart2}
\G_i^{(2)}|_{\rm{II}}&=& 4 g^2\sum_{i\neq j\neq k}\widehat B_1(q_i-q_j)\widehat B_3(q_i-q_k)\nonumber\\
&=&2 g^2\sum_{i\neq j\neq k} \bigg( \widehat B_1(q_i-q_j)\widehat B_3(q_i-q_k)+ \widehat B_1(q_i-q_k)\widehat B_3(q_i-q_j) \bigg)\,.
\ea

The next step is to rewrite the simplified two-loop effective potential in (\ref{sungfpart1}), (\ref{sungfpart2}), (\ref{sungipart1}) and (\ref{sungipart2}) in terms of $\widehat B_4(x)$.  Firstly, we consider $\Gamma_{f}^{(2)}|_{\rm{I}}$ and $\Gamma_{i}^{(2)}|_{\rm{I}}$. These two terms are easy to deal with because there is only one argument $q_i - q_j$ in the Bernoulli polynomials.

Using the definition of the Bernoulli polynomials, we can show the following two identities
\ba
\la{id1}
(\widehat B_2(q_i - q_j)-\widehat B_2(0))^2&=&\frac{3}{8 \pi^2}
\widehat B_4(q_i - q_j) \, ,\\
\la{id2}
\widehat B_2(0) \widehat B_2(q_i-q_j) + \widehat B_1(q_i-q_j)\widehat B_3(q_i-q_j)&=&\frac{T^4}{144}-\frac{1}{4 \pi^2} \widehat B_4 (q_i-q_j)\, .
\ea
Using (\ref{id1}), we can express $\Gamma_{f}^{(2)}|_{\rm{I}}$ as
\ba
\Gamma_{f}^{(2)}|_{\rm{I}}&=&g^2\sum_{i\neq j} \bigg(4 \widehat B_2(0) \widehat B_2(q_i - q_j) + \frac{3}{8 \pi^2}
\widehat B_4(q_i - q_j) -(\widehat B_2(0))^2 \bigg)\,,\nonumber\\&=&4 g^2\sum_{i\neq j}  \widehat B_2(0) \widehat B_2(q_i - q_j) + \frac{3 g^2}{8 \pi^2} \sum_{i\neq j}
\widehat B_4(q_i - q_j) - \frac{g^2 T^4}{144} N(N-1)\, .
\ea
Furthermore, by using (\ref{id2}), the sum of $\Gamma_{f}^{(2)}|_{\rm{I}}$ and $\Gamma_{i}^{(2)}|_{\rm{I}}$ becomes
\be
\la{firsttype}
\Gamma^{(2)}|_{\rm{I}} \equiv \Gamma_{f}^{(2)}|_{\rm{I}}+\Gamma_{i}^{(2)}|_{\rm{I}}=-\frac{5 g^2 }{8 \pi^2} \sum_{i\neq j}
\widehat B_4(q_i - q_j) + \frac{3 g^2 T^4}{144} N(N-1)\, .
\ee
It is actually the final result we need for $\Gamma^{(2)}|_{\rm{I}}$.

However, for $\Gamma_{f}^{(2)}|_{\rm{II}}$ and $\Gamma_{i}^{(2)}|_{\rm{II}}$, we have two different arguments $q_i - q_j$ and $q_i - q_k$ in the Bernoulli polynomials simultaneously. To complete this proof, it is very important to use the following non-trivial identity
\be
\la{id3}
{\cal F}(q_i,q_j,q_k)+{\cal F}(q_j,q_k,q_i)+{\cal F}(q_k,q_i,q_j)=\frac{T^4}{48}-\frac{5}{16 \pi^2}(\widehat B_4(q_i - q_j)+\widehat B_4(q_i - q_k)+\widehat B_4(q_j - q_k))\, ,
\ee
with
\be
\la{fijk}
{\cal F}(q_i,q_j,q_k) \equiv 2 \widehat B_1(q_i-q_j)\widehat B_3(q_i-q_k)+2\widehat  B_1(q_i-q_k)\widehat B_3(q_i-q_j)+ \widehat B_2(q_i-q_j)\widehat B_2(q_i-q_k)\,.
\ee

The proof of (\ref{id3}) is given in Appendix \ref{proofid3}. After this identity is proved, it is straightforward to get the following result
\ba
\la{secondtype}
\Gamma_{f}^{(2)}|_{\rm{II}}+\Gamma_{i}^{(2)}|_{\rm{II}}&=&
g^2\sum_{i\neq j\neq k} {\cal F}(q_i,q_j,q_k)\nonumber \\
&=&g^2\sum_{i\neq j\neq k} \frac{1}{3} \bigg(\frac{T^4}{48}-\frac{5}{16 \pi^2}(\widehat B_4(q_i-q_j)+\widehat B_4(q_i-q_k)+ \widehat B_4(q_j-q_k))\bigg)\nonumber \\
&=& \frac{g^2 T^4}{144}N(N-1)(N-2)-\frac{5 g^2}{16 \pi^2}(N-2)\sum_{i\neq j}\widehat B_4(q_i-q_j)\, .
\ea
In the third and second lines of (\ref{secondtype}), we have used the following two equations, respectively.
\ba
\sum_{i\neq j\neq k} B_4(q_i-q_j)&=&\sum_{i\neq j\neq k} \widehat B_4(q_i-q_k)=\sum_{i\neq j\neq k}  \widehat B_4(q_j-q_k)=\sum_{i\neq j} (N-2)\widehat B_4(q_i-q_j)\, ,\nonumber \\
\sum_{i\neq j\neq k} {\cal F}_{ijk}&=&\sum_{i\neq j\neq k} {\cal F}_{jki}=\sum_{i\neq j\neq k} {\cal F}_{kij}\, .
\ea
These equations can be obtained by using the fact that $i\neq j\neq k$ indicates the three indices $i$, $j$ and $k$ are equivalent which can be interchanged under the summation.

Combine (\ref{firsttype}) and (\ref{secondtype}) and use the expression of the one-loop effective potential
\be
\la{sun1loop}
\Gamma^{(1)}= - \frac{\pi^2 T^4}{45} (N^2-1)+ \sum_{i\neq j} \widehat B_4(q_i - q_j) \, ,
\ee
we finally prove the simple relation between one- and two-loop effective potential for $SU(N)$ as given in (\ref{rela}).

\subsection{Proof for $SO(2N)$ and $SO(2N+1)$}

%For these groups we use a variant of the notation from Georgi's
%book~\cite{georgi}. The generators $M^{ab}$ in the fundamental
%representation have matrix elements,
%
%\be
%(M^{ab})_{xy}=-\frac{i}{2}(\d_{ax}\d_{by}-\d_{ay}\d_{bx})\, .
%\ee
%
%Obviously there is antisymmetry under exchange of the labels $a$ and
%$b$, i.e.\ $M^{ab}=-M^{ba}$.

%Furthermore, we can define the off-diagonal generators in the Cartan
%basis. For both groups, there are $N(2N-2)$ off-diagonal generators
%$E^{\eta i. \eta^\prime j}$ with $i,j=1,\cdots,N$ and $i>j$. Here, we
%define the indices $i$ with an associated sign $\eta$. Similarly, $j$
%is defined with $\eta^\prime$. The signs $\eta$ or $\eta^\prime$ are
%independently $\pm 1$. The explicit form of the generators is
%
%\be E^{\eta i. \eta^\prime j}=\frac{1}{2}(M^{2i-1,2j-1}+i \eta
%M^{2i,2j-1}+i \eta^\prime M^{2i-1,2j}-\eta \eta^\prime M^{2i,2j})\,.
%\ee
%
%For $SO(2N+1)$ there are $2N$ additional off-diagonal generators
%
%\be
%E^{\eta i}=\frac{1}{\sqrt{2}}(M^{2i-1,2N+1}+i \eta M^{2i,2N+1})\,.
%\ee
%
%For either of the groups the $N$-dimensional Cartan subalgebra is spanned by
%mutually commuting and orthogonal generators $H^d$, with
%
%\be
%H^d=M^{2d-1,2d}\, ,\, {\rm with} \,\,\,d=1,2,\cdots\,N.
%\la{eq:diagcartan}
%\ee

For both groups, there are $N(2N-2)$ off-diagonal generators
$E^{\eta i. \eta^\prime j}$,
\be E^{\eta i. \eta^\prime j}=\frac{1}{2}(M^{2i-1,2j-1}+i \eta
M^{2i,2j-1}+i \eta^\prime M^{2i-1,2j}-\eta \eta^\prime M^{2i,2j})\,,
\ee
where
\be
(M^{ab})_{xy}=-\frac{i}{2}(\d_{ax}\d_{by}-\d_{ay}\d_{bx})\, .
\ee
In the above equation, $i,j=1,\cdots,N$ and $i>j$. We
define the indices $i$ with an associated sign $\eta$. Similarly, $j$
is defined with $\eta^\prime$. The signs $\eta$ or $\eta^\prime$ are
independently $\pm 1$.

For $SO(2N+1)$ there are $2N$ additional off-diagonal generators
\be
E^{\eta i}=\frac{1}{\sqrt{2}}(M^{2i-1,2N+1}+i \eta M^{2i,2N+1})\,.
\ee

For either of the groups the $N$-dimensional Cartan subalgebra is spanned by
mutually commuting and orthogonal generators $H^d$, with
\be
H^d=M^{2d-1,2d}\, ,\, {\rm with} \,\,\,d=1,2,\cdots\,N.
\la{eq:diagcartan}
\ee

The structure constants can be obtained from the commutation
relations\footnote{For $SO(2N)$ and $SO(2N+1)$, if a typical
  off-diagonal index $b$ is denoted as $b=\eta i.\eta^\prime j$ then
  $-b=-\eta i.-\eta^\prime j$; if $b=\eta i$, then $-b=-\eta i$. In
  (\ref{4com}), with our notation, $E^{\rho k.\eta i}$ should be
  understood as $-E^{\eta i.\rho k}$ if $i>k$. Similarly for
  $E^{\eta^\prime j.\rho^\prime l}$.}
\ba \la{1com} [H^d,E^{\eta j}]&\equiv& f^{d,\eta j,-\eta^\prime k}
E^{\eta^\prime k}=\frac{\eta}{2}\d_{dj} E^{\eta j}\\{}\la{2com}
[H^d,E^{\eta j.\eta^\prime k}]&\equiv& f^{d,\eta j.\eta^\prime k
  ,-\rho l.-\rho^\prime m} E^{\rho l.\rho^\prime
  m}=\frac{1}{2}(\eta\d_{dj}+\eta^\prime\d_{dk}) E^{\eta j.\eta^\prime
  k}\\{}\la{3com} [E^{\eta i.\eta^\prime j},E^{\rho k}]&\equiv&
f^{\eta i.\eta^\prime j, \rho k, -\sigma l} E^{\sigma
  l}=\frac{i}{4}\bigg(\d_{ki}(1-\rho \eta)E^{\eta^\prime
  j}-\d_{kj}(1-\rho \eta^\prime)E^{\eta i}\bigg)\\{} [E^{\eta
    i.\eta^\prime j},E^{\rho k.\rho^\prime l}]&\equiv& f^{\eta
  i.\eta^\prime j,\rho k.\rho^\prime l,-\sigma t.-\sigma^\prime n}
E^{\sigma t.\sigma^\prime n}=\frac{i}{4}\bigg(\d_{ki}(1-\rho
\eta)E^{\eta^\prime j.\rho^\prime l} -\nonumber \\
&&\hspace{-1cm} \d_{kj} (1-\rho
\eta^\prime)E^{\eta i.\rho^\prime l}-\d_{lj}(1-\rho^\prime
\eta^\prime)E^{\rho k.\eta i}+\d_{il}(1-\eta \rho^\prime)E^{\rho
  k.\eta^\prime j}\bigg)\,. \la{4com} \ea
From (\ref{1com}) and (\ref{2com}), the roots can be expressed as
\ba
\la{r1}
{\vec \a}^{\,\eta i}&=&\frac{\eta}{2}{\vec e}_i\, , \\
\la{r2}
{\vec \a}^{\,\eta i.\eta^\prime j}&=&\frac{1}{2}(\eta{\vec e}_i+\eta^\prime{\vec e}_j)\, .
\ea
From (\ref{3com}) and (\ref{4com}), the non-vanishing structure constants are given by
\ba
\la{sc1}
f^{\,\eta i.\eta^\prime j,-\eta i,-\eta^\prime j}&=&\frac{i}{2}\, ,\quad {\rm with}\quad i>j\, ,  \\
\la{sc2}
f^{\,\eta i.\eta^\prime j,-\eta i.\rho l,-\eta^\prime j. -\rho l}&=&\frac{i}{2}\, ,\quad {\rm with}\quad i>j>l\,.
\ea
%There are $N(2N-2)$ off-diagonal generators associated with the long
%roots and $2N$ off-diagonal generators associated with the short
%roots. For both $SO(2N)$ and $SO(2N+1)$, $d(r)=N$. Using
%Eq.~(\ref{casimirroots}) we can easily get $C_2(A)=N-\frac{1}{2}$ for
%$SO(2N+1)$ and $C_2(A)=N-1$ for $SO(2N)$.

For these two groups, the arguments in the Bernoulli functions have three different cases: for a diagonal index $a$, $q_a = 0$; for an off-diagonal index $a=\eta i$, $q_a = \eta q_i$ and for an off-diagonal index $a=\eta i.\eta^\prime j$, $q_a = \eta q_i+\eta^\prime q_j$.

We start with $SO(2N)$ and split the two-loop effective potential into two parts. For $\Gamma_{f}^{(2)}$, we have
\ba
\Gamma_{f}^{(2)}|_{\rm{I}} &=& \frac{g^2}{2}\sum_{i\neq j} \bigg(2 \widehat B_2(0) \widehat B_2(q_i \pm q_j) +
(\widehat B_2(q_i + q_j))^2+
(\widehat B_2(q_i - q_j))^2\bigg)\, , \\
\la{gammaf2}
\Gamma_{f}^{(2)}|_{\rm{II}} &=& \frac{g^2}{2}\sum_{i\neq j \neq k}  \widehat B_2(q_i \pm q_j)
\widehat B_2(q_i \pm q_k)\, .
\ea

In order to get (\ref{gammaf2}), we should mention that there are six terms which contribute to the final result and they correspond to six possible ways to get a non-vanishing structure constant. If we symbolically write the structure constant in (\ref{sc2}) as $f^{AB,CD,EF}$, then the six possible ways are the following\footnote{With this notation, if $A=\eta i$ and $B= \eta' j$, then $A=B$ means $\eta=-\eta'$, $i=j$.}:
\ba
&&A=C,B=E,D=F\,; \quad A=F,B=D,C=E\,; \quad A=C,B=F,D=E\,;\nonumber\\
&&A=D,B=F,C=E\,; \quad A=E,B=C,D=F\,; \quad A=E,B=D,C=F\,.
\ea
In fact, every term has exactly the same form as that in (\ref{gammaf2}) expect for the constraint in the summation. The constraint for an individual term is just one possible way to order the values of $i$, $j$ and $k$, for example $i>j>k$. As a result, the total contribution is given by (\ref{gammaf2}) with the constraint ${i\neq j \neq k}$.

Similarly as $SU(N)$$, \Gamma_{f}^{(2)}|_{\rm{I}}$ corresponds to the case where only one index is diagonal in (\ref{newuf}) while $\Gamma_{f}^{(2)}|_{\rm{II}}$ is the part with all the indices being off-diagonal in (\ref{newuf}). To get these two equations, we have used (\ref{r2}) and (\ref{sc2}) to get rid of the structure constants. To keep the notations compact, we define $\widehat B_n(q_i \pm q_j)\equiv \widehat B_n(q_i + q_j)+\widehat B_n(q_i - q_j)$ with $n=1,2,3,4$.

For the $\Gamma_{i}^{(2)}$, we have
\ba
\la{gi21}
\Gamma_{i}^{(2)}|_{\rm{I}} &=& 2 g^2 \sum_{i\neq j} \widehat B_1(q_i \pm q_j)
\widehat B_3(q_i \pm q_j)\, , \nonumber\\
&=&2 g^2 \sum_{i\neq j}\bigg( \widehat B_1(q_i + q_j)\widehat B_3(q_i + q_j)+\widehat B_1(q_i - q_j)
\widehat B_3(q_i - q_j)\bigg)\, ,\\
\la{gi22}
\Gamma_{i}^{(2)}|_{\rm{II}} &=& g^2 \sum_{i\neq j \neq k}  \widehat B_1(q_i \pm q_j)
\widehat B_3(q_i \pm q_k) + \widehat B_1(q_i \pm q_k)
\widehat B_3(q_i \pm q_j)\, .
\ea
In the second line of (\ref{gi21}), we use the fact that $\sum_{i\neq j}\widehat B_1(q_i + q_j)\widehat B_3(q_i - q_j)=\sum_{i\neq j}\widehat B_1(q_i - q_j)\widehat B_3(q_i + q_j)=0$ because $\widehat B_1(x)$ and $\widehat B_3(x)$ are odd functions of $x$.

For the sum of $\Gamma_{i}^{(2)}|_{\rm{I}}$ and $\Gamma_{f}^{(2)}|_{\rm{I}}$, by using (\ref{id1}) and (\ref{id2}), it is easy to show that
\be
\la{so2npart1}
\Gamma_{i}^{(2)}|_{\rm{I}}+\Gamma_{f}^{(2)}|_{\rm{I}}=\frac{-5g^2}{16 \pi^2} \sum_{i\neq j} \widehat B_4(q_i \pm q_j) +\frac{3g^2 T^3}{144} (N^2-N)\, .
\ee

Before we discuss $\Gamma_{i}^{(2)}|_{\rm{II}}$ and $\Gamma_{f}^{(2)}|_{\rm{II}}$, we need to prove the following identity
\be
\la{id4}
{\cal G}(q_i,q_j,q_k)+{\cal G}(q_j,q_k,q_i)+{\cal G}(q_k,q_i,q_j)=\frac{T^4}{12}-\frac{5}{8 \pi^2}(\widehat B_4(q_i \pm q_j)+\widehat B_4(q_i \pm q_k)+\widehat B_4(q_j \pm q_k))\, ,
\ee
with
\be
\la{fijk}
{\cal G}(q_i,q_j,q_k) \equiv 2 \widehat B_1(q_i\pm q_j)\widehat B_3(q_i \pm q_k)+ 2 \widehat  B_1(q_i \pm q_k)\widehat B_3(q_i\pm q_j)+ \widehat B_2(q_i\pm q_j)\widehat B_2(q_i\pm q_k)\,.
\ee
Based on the identity given in (\ref{id3}), we can easily prove (\ref{id4}) by using the following relation
\be
{\cal G}(q_i,q_j,q_k)={\cal F}(q_i,q_j,q_k)+{\cal F}(q_i,-q_j,-q_k)+{\cal F}(q_i,-q_j,q_k)+{\cal F}(q_i,q_j,-q_k)\, .
\ee

Then the sum of $\Gamma_{i}^{(2)}|_{\rm{II}}$ and $\Gamma_{f}^{(2)}|_{\rm{II}}$ can be obtained by summing over the indices $i$, $j$ and $k$ in (\ref{id4}) with the condition $i\neq j \neq k$.
\ba
\la{so2npart2}
\Gamma_{i}^{(2)}|_{\rm{II}}+\Gamma_{f}^{(2)}|_{\rm{II}}&=&\frac{g^2}{2} \sum_{i\neq j \neq k}{\cal G}(q_i,q_j,q_k)\nonumber \\ &=& \frac{g^2 T^4}{72} (N^2-N)(N-2)-\frac{5g^2}{16 \pi^2}(N-2) \sum_{i\neq j} \widehat B_4(q_i \pm q_j) \, .
\ea

Adding up (\ref{so2npart1}) and (\ref{so2npart2}) and using the expression of the one-loop effective potential
\be
\la{so21loop}
\Gamma^{(1)}= - \frac{\pi^2 T^4}{45} N (2N-1)+ \sum_{i\neq j} \widehat B_4(q_i \pm q_j) \, ,
\ee
we complete the proof of (\ref{rela}) for $SO(2N)$.

Once $SO(2N)$ is done, the proof for $SO(2N+1)$ is easy. We only need to consider the extra terms in the effective potential which arise due to the $2N$ off-diagonal generators $E^{\eta i}$. Using (\ref{r1}) and (\ref{sc1}), we can get
\ba
\la{gf21so2n1}
\Gamma_{f}^{(2)}|_{\rm{I}} &=& \frac{g^2}{2}\sum_{i} \bigg(2 \widehat B_2(0) \widehat B_2(q_i) +
(\widehat B_2(q_i))^2\bigg)\, , \\
\Gamma_{f}^{(2)}|_{\rm{II}} &=& g^2 \sum_{i\neq j} \bigg(\widehat B_2(q_i)
\widehat B_2(q_j)+\frac{1}{2}\widehat B_2(q_i \pm q_j)\widehat B_2(q_i)+\frac{1}{2}\widehat B_2(q_j \pm q_i)\widehat B_2(q_j)\bigg)\, .
\ea

For the $\Gamma_{i}^{(2)}$, we have
\ba
\la{gi21so2n1}
\Gamma_{i}^{(2)}|_{\rm{I}} &=& 2 g^2 \sum_{i} \widehat B_1(q_i)
\widehat B_3(q_i)\, , \\
\Gamma_{i}^{(2)}|_{\rm{II}} &=& g^2 \sum_{i\neq j} \bigg( \widehat B_1(q_i \pm q_j)
\widehat B_3(q_i) + \widehat B_1(q_i)
\widehat B_3(q_i \pm q_j)\nonumber \\&+&\widehat B_1(q_j \pm q_i)
\widehat B_3(q_j) + \widehat B_1(q_j)
\widehat B_3(q_j \pm q_i)\bigg)\, .
\ea

Combing (\ref{gf21so2n1}) and (\ref{gi21so2n1}) and using (\ref{id1}) and (\ref{id2}), we can show
\be
\la{so2n+1part1}
\Gamma_{f}^{(2)}|_{\rm{I}} +\Gamma_{i}^{(2)}|_{\rm{I}} = \frac{g^2 T^4}{96} N-\frac{5 g^2}{16 \pi^2}\sum_{i} \widehat B_4(q_i)\, .
\ee

In (\ref{id4}), by setting $q_k=0$, the sum of $\Gamma_{f}^{(2)}|_{\rm{II}}$ and $\Gamma_{i}^{(2)}|_{\rm{II}}$ then has the following form
\ba
\la{so2n+1part2}
&&\Gamma_{f}^{(2)}|_{\rm{II}} +\Gamma_{i}^{(2)}|_{\rm{II}}= \frac{g^2}{4}\sum_{i\neq j}\bigg({\cal G}(q_i,q_j,0)+ {\cal G}(q_j,0,q_i)+{\cal G}(0,q_i,q_j)\bigg)\nonumber \\ &=& g^2 \sum_{i\neq j} \bigg(\frac{T^4}{48}-\frac{5}{32 \pi^2}(\widehat B_4(q_i \pm q_j)+2 \widehat B_4(q_i)+2\widehat B_4(q_j ))\bigg)\nonumber\\ &=& \frac{g^2 T^4}{48}(N^2-N)-\frac{5 g^2}{16 \pi^2}(2N-2) \sum_{i}\widehat B_4(q_i)-\frac{5 g^2}{32 \pi^2}\sum_{i\neq j}\widehat B_4(q_i \pm q_j)\, .
\ea

The one-loop effective potential for $SO(2N+1)$ is given by
\be
\la{so2n+11loop}
\Gamma^{(1)}= - \frac{\pi^2 T^4}{45} N (2N+1)+ \sum_{i\neq j} \widehat B_4(q_i \pm q_j)+2 \sum_{i} \widehat B_4(q_i) \, .
\ee

Adding up (\ref{so2npart1}), (\ref{so2npart2}), (\ref{so2n+1part1}) and (\ref{so2n+1part2}), we prove (\ref{rela}) for $SO(2N+1)$.

\subsection{Proof for $Sp(2N)$}

%In this section we discuss the symplectic groups $Sp(2N)$. They are
%the pseudoreal part of $SU(2N)$ constructed by defining the charge
%conjugation matrix,
%
%\be
%  I_{2N}=i\s_2\otimes {\bf 1}_N ~, \la{eq:ccmatrix}
%\ee
%
%and requiring the special unitary matrix $U$ to obey
%
%\be I_{2N}\; U\;
%I_{2N}^\dagger=U^* ~,  \la{eq:symplectic}
%\ee
%
%where $\s_i$ are the Pauli matrices with $i=1,2,3$, and ${\bf 1}_N$ is
%the $N$-dimensional unit matrix.

%Writing $U={\rm exp}(i {\cal G})$, the symplectic generator is of the form
%
%\ba
%{\cal G}=\left (\begin{array}{cc} A&B\\
%B^*&-A^*
%\end{array}\right)
%\label{eq:symgen}\, .
%\ea
%
%Here, $A$ is a Hermitian matrix with $A=A^\dagger$, and $B=B^t$ is
%complex. For $N=1$, this form indeed reduces to the generator of
%$SU(2)$. The Hermitian matrix A is not traceless, but ${\cal G}$
%is. We therefore have $N^2$ real degrees of freedom from $A$, and
%$N(N+1)$ degrees of freedom from the symmetric complex matrix $B$. In
%total, we have $N(2N+1)$. The Cartan space is $N$ dimensional.

The diagonal generators of $Sp(2N)$ is
\ba
H^d={1\over{\sqrt{2}}}\; \s_3\otimes \l^d\,,~d=1,\cdots,N ~.
\ea
Here, $\s_i$ are the Pauli matrices with $i=1,2,3$. The $N-1$ matrices $\l^d$ are the same as for $SU(N)$, and
we need the additional $\l^N={1\over{\sqrt{2N}}}{\bf 1}_N$. ${\bf 1}_N$ is
the $N$-dimensional unit matrix.

The corresponding off-diagonal generators $E^{ij}$ are
\ba
E^{ij}={1\over{\sqrt{2}}}\left (\begin{array}{cc} \l^{ij}&0\\
0&-\l^{ji}
\end{array}\right)%,~\mbox{or simply}\nonumber\\
%H_d&=&\{1\over{\sqrt{2}}}s_3\otimes \l_d
\label{eq:symgendiag},~i,j=1,\cdots,N\,,\quad {\rm and}\quad i\neq j\,.
\ea

In addition, we have additional $N(N+1)$ off-diagonal generators denoted as $E^{\eta ij}$; the
first index $\eta$ is a sign index. They are defined by
\ba
\la{Eije}
E^{\eta ij} &=&
\left[
  {1\over{\sqrt{2}}}+\d_{ij}(\frac{1}{2}-\frac{1}{\sqrt{2}})\right]
\; \s^{\eta}\otimes
(\l^{ij}+\l^{ji})\,, ~i,j=1,\cdots,N,\quad {\rm and}\quad i\ge j\,,
\ea
where $\s^{\eta}={1\over 2}(\s_1+ i \eta \s_2)$.

The structure constants can be obtained from the commutation relations\footnote{For $Sp(2N)$, if a typical off-diagonal index $b$ is denoted as $b = ij$, then $- b = ji$ ; if $b = \eta ij$, then $-b = - \eta ij$.}
\ba
~[H^{d},E^{ij}]&=&\frac{1}{\sqrt{2}}(\l^d_{ii}-\l^d_{jj})E^{ij}\,, \\
~[H^{d},E^{\eta ij}]&=&\frac{\eta}{\sqrt{2}}(\l^d_{ii}+\l^d_{jj})E^{\eta ij}\quad {\rm for} \quad i>j \,, \\
~[H^{d},E^{\eta ii}]&=&\eta\sqrt{2}\l^d_{ii}E^{\eta ii}\quad {\rm for} \quad i=j \,, \\
\la{spcr1}
~[E^{ij},E^{kl}]&=&\frac{1}{2}(\d_{jk}E^{il}-\d_{il}E^{kj})\,,\\
\la{spcr2}
~[E^{+ij},E^{-kl}]&=&{1\over 2}(\d_{jk}E^{il}+\d_{il}E^{jk}+\d_{jl}E^{ik}+\d_{ik}E^{jl})\quad {\rm for} \quad i \neq j \,\, {\rm and}\,\, k \neq l\,,\\
\la{spcr3}
~[E^{+ii},E^{-kl}]&=&{1\over \sqrt{2}}(\d_{ik}E^{il}+\d_{il}E^{ik})\quad {\rm for} \quad k \neq l\,,\\
\la{spcr4}
~[E^{+kl},E^{-ii}]&=&{1\over \sqrt{2}}(\d_{ik}E^{li}+\d_{il}E^{ki})\quad {\rm for} \quad k \neq l\,,\\
~[E^{\eta ij},E^{\eta kl}]&=&0\,.
\ea

Like for $SU(N)$, we can write the roots for $Sp(2N)$
in terms of the orthonormal basis $\{{\vec e}_i\}$,
\ba
\vec\a^{\,\eta ij}&=& {\eta\over 2}( \vec e_i +\vec e_j)\,,~1\le j <
i\le N\, ,\nonumber\\
\vec\a^{\,ij}&=&{1\over 2}(\vec e_i-\vec e_j)\, ,~1\le i\le N\, ,~1\le
j\le N\, ~{\rm and}\,~ i\neq j\,,\nonumber\\
\vec\a^{\,\eta i}&=&\eta \vec e_i,~1\le i\le N\,.
\la{eq:rootsspninebasis}
\ea
Here, the first roots are associated with the generators $E^{\eta ij}$
when $i>j$ and the second roots are associated with the generators
$E^{ij}$.  The roots
$\vec\a^{\,\eta i}$ (which can be also written as $\vec\a^{\,\eta
  ii}$) come from $E^{\eta ii}$.

For $Sp(2N)$, we found that the arguments of the Bernoulli functions are the following: for a diagonal index, $q_d = 0$; for an off-diagonal index, we have $q_{ij} = q_i - q_j$ and $q_{\eta ij} = \eta (q_i + q_j)$.

Using the results given in (\ref{eq:rootsspninebasis}), the contribution with one diagonal index in the structure constant in (\ref{newuf}) is given by
\ba
\Gamma_{f}^{(2)}|_{\rm{I}} &=& \frac{g^2}{2}\sum_{i\neq j} \bigg(2 \widehat B_2(0) \widehat B_2(q_i - q_j) +
(\widehat B_2(q_i - q_j))^2+2 \widehat B_2(0) \widehat B_2(q_i + q_j)\nonumber \\ &+&
(\widehat B_2(q_i + q_j))^2\bigg)+ g^2\sum_{i} \bigg(4 \widehat B_2(0) \widehat B_2(2 q_i ) + 2
(\widehat B_2(2 q_i ))^2\bigg)\, .
\ea
For other contributions from (\ref{newuf}), we have
\ba
\la{spgf2}
\Gamma_{f}^{(2)}|_{\rm{II}} &=& \frac{g^2}{2}\sum_{i\neq j\neq k} \bigg( \widehat B_2(q_i -q_j) \widehat B_2(q_i - q_k) \nonumber \\ &+&
\widehat B_2(q_i + q_j)\widehat B_2(q_i + q_k)+ 2 \widehat B_2(q_i-q_k) \widehat B_2(q_i + q_j)\bigg)\nonumber \\ &+& 2 g^2 \sum_{i\neq j}\bigg(
\widehat B_2(2 q_i) \widehat B_2(q_i \pm q_j)+\widehat B_2(q_i+q_j) \widehat B_2(q_i -q_j )\bigg)\, .
\ea
In the above equation, the first line is obtained by using (\ref{spcr1}) which is the same as the corresponding contribution in $SU(N)$ up to a constant factor; to derive the second line in (\ref{spgf2}), we used (\ref{spcr2}) which gives the non-vanishing structure constants as the following
\ba
f^{+ij,-jk,ki}_{i>j>k}=f^{+ij,-kj,ki}_{i>k>j}=f^{+ji,-kj,ki}_{k>j>i}=f^{+ij,-kj,ki}_{k>i>j}
=f^{+ji,-jk,ki}_{j>i>k}=f^{+ji,-jk,ki}_{j>k>i}=\frac{1}{2}\, .
\ea
Here, the subscript indicates the constraint on the indices for each structure constant. It is easy to check that each contribution related to the individual structure constant as given in the above equation has the same form and sum of the six terms corresponds to the constraint $i \neq j \neq k$ in (\ref{spgf2}).

Similarly, we can get the following non-vanishing structure constants from Eqs.~(\ref{spcr3}) and (\ref{spcr4})
\ba
f^{+ii,-ij,ji}=f^{+ii,-ji,ji}=f^{+ij,-ii,ij}=f^{+ji,-ii,ij}=\frac{1}{\sqrt{2}}\, .
\ea
Using this explicit result, the last line of (\ref{spgf2}) is obtained.

The contribution from the insertion diagram given in (\ref{g2sp2nsim}) has to be rewritten in a proper way in order to combine with the corresponding terms in $\Gamma_f^{(2)}$. As before, we split it into two parts
\ba
\la{g2sp2nsim}
\G_i^{(2)}|_{\rm I}&=&2 g^2\sum_{i\neq j}\bigg(\widehat
B_1(q_i+q_j)\widehat B_3(q_i+q_j)+\widehat B_1(q_i-q_j)\widehat
B_3(q_i-q_j)\bigg) \nonumber \\&+& 8 g^2\sum_{i}\widehat
B_1(2 q_i)\widehat B_3(2 q_i)\, ,\nonumber \\
\G_i^{(2)}|_{\rm II}&=&2 g^2\sum_{i\neq j \neq k}\widehat
B_1(q_i\pm q_j)\widehat B_3(q_i\pm q_k)+ 4 g^2\sum_{i\neq j}\bigg(\widehat
B_1(q_i\pm q_j)\widehat B_3(2 q_i)\nonumber \\&+&\widehat B_1(2 q_i)\widehat
B_3(q_i\pm q_j)\bigg)\, .
\ea
Similarly as what we did in (\ref{gi21}), terms which have zero contribution have been dropped in $\G_i^{(2)}$.

At this point, we can easily show that the combination of $\G_f^{(2)}|_{\rm I}$ and $\G_i^{(2)}|_{\rm I}$ can be expressed in terms of $\widehat B_4(x)$ as
\ba
\la{sppart1}
\G_f^{(2)}|_{\rm I}+\G_i^{(2)}|_{\rm I}=\frac{g^2 T^4}{48}N(N+1)-\frac{5 g^2}{16 \pi^2}\sum_{i\neq j}\widehat B_4(q_i\pm q_j)-\frac{5 g^2}{4 \pi^2}\sum_{i}\widehat B_4(2 q_i)
\ea

For the contributions from $\G_f^{(2)}|_{\rm II}$ and $\G_i^{(2)}|_{\rm II}$, we can rewrite the result in a proper way  in order to make use of (\ref{id4}),
\ba
\la{newspg12}
\G_f^{(2)}|_{\rm II}+\G_i^{(2)}|_{\rm II}&=& \frac{g^2}{2}\sum_{i\neq j \neq k}\bigg(\widehat B_2(q_i\pm q_j) \widehat B_2(q_i \pm q_k)+2 \widehat
B_1(q_i\pm q_j)\widehat B_3(q_i\pm q_k)\nonumber \\&+&2 \widehat
B_1(q_i\pm q_k)\widehat B_3(q_i\pm q_j)\bigg)+ 2 g^2\sum_{i\neq j}\bigg(\widehat B_2(2 q_i) \widehat B_2(q_i \pm q_j)\nonumber \\&+&\widehat B_2(q_i+q_j) \widehat B_2(q_i -q_j )+2\widehat
B_1(q_i\pm q_j)\widehat B_3(2 q_i)+2\widehat B_1(2 q_i)\nonumber \\&&\widehat
B_3(q_i\pm q_j)\bigg)\, .
\ea

The first summation in the above equation is the same as $SO(2N)$ and the result is given in (\ref{so2npart2}). In order to perform the second summation, we can use (\ref{id4}) in the case $q_i=q_k$ which gives
\ba
\la{spsumpart2}
&&2 \widehat B_2(2q_i) \widehat B_2(q_i \pm q_j)+2 \widehat B_2(q_i + q_j)\widehat B_2(q_i - q_j)\nonumber \\&&+4\widehat B_1(q_i \pm q_j)\widehat B_3(2 q_i)+4\widehat B_1(2 q_i)\widehat B_3(q_i \pm q_j)-\frac{5}{8\pi^2}\widehat B_4(q_i \pm q_j)+\frac{T^4}{24}\nonumber \\&&+4\widehat B_1(q_j + q_i)\widehat B_3(q_j - q_i)+4\widehat B_1(q_j - q_i)\widehat B_3(q_j + q_i) \nonumber \\&&=\frac{T^4}{12}-\frac{5}{8\pi^2}(2\widehat B_4(q_i \pm q_j)+\widehat B_4(2 q_i))\, .
\ea
Here, we have used (\ref{id1}) and (\ref{id2}) and some vanishing terms are dropped according to (\ref{b2can}). Then sum over indices $i$ and $j$ with $i\neq j$ in (\ref{spsumpart2}), we find the result for the second summation in (\ref{newspg12}) reads
\be
\la{secondsum}
-\frac{5g^2}{8\pi^2}\sum_{i\neq j}\widehat B_4(q_i \pm q_j)-\frac{5g^2}{8\pi^2}(N-1)\sum_{i}\widehat B_4(2q_i)+\frac{g^2 T^4}{24}N(N-1)\, .
\ee
Notice that the third line in (\ref{spsumpart2}) vanishes under the summation.

Now we can write down the final result for (\ref{newspg12})
\be
\la{spg12result}
\G_f^{(2)}|_{\rm II}+\G_i^{(2)}|_{\rm II}= -\frac{5g^2}{16\pi^2}N \sum_{i\neq j}\widehat B_4(q_i \pm q_j)-\frac{5g^2}{8\pi^2}(N-1)\sum_{i}\widehat B_4(2q_i)+\frac{g^2 T^4}{72}(N^2-N)(N+1)\, .
\ee
Together with (\ref{sppart1}) and the one-loop effective potential for $Sp(2N)$
\be
\la{sp2n1loop}
\Gamma^{(1)}= - \frac{\pi^2 T^4}{45} N (2N+1)+ \sum_{i\neq j} \widehat B_4(q_i \pm q_j)+ 2 \sum_{i} \widehat B_4(2 q_i) \, ,
\ee
we prove (\ref{rela}) for $Sp(2N)$.

\section{Conclusions and Outlook}\la{sec:conclusions}

We reviewed the basic ideas of the construction of the matrix model which were first proposed by Meisinger, Miller and Ogilvie. The non-ideal contributions in this model have been further improved for a quantitative fit to the lattice simulations on the thermodynamics for $SU(N)$ gauge theories. Previous work included the Bernoulli Polynomial $\widehat B_4(x)$ in the non-ideal contributions for such a purpose which reproduced the lattice result in the semi-QGP region very well. In this work, we consider the temperature dependence of the mass scale appears in the original matrix model, and the corresponding new model also does a good job in the semi-QGP region, therefore, can be treated as an alternative to the previous models. On the other hand, with the exact temperature dependence given in (\ref{mass}), our new model is able to get the upward trend of the rescaled trace anomaly as found by the recent lattice simulations. Starting at about $4T_c$, this quantity increases with temperature and such a behavior can not be obtained by using previous matrix models. Up to $8T_c$, the new model is in good agreement with lattice results. Beyond this temperature, the HTL perturbation theory works well.

In addition, we discussed the perturbative corrections to the one-loop thermal effective potential which is used as the ideal contributions for all the matrix models. In particular, there is a simple relation between the one- and two-loop effective potential as shown in (\ref{rela}). An analytical proof of this relation is first given in this paper. We show it is quite general and holds for all the classic groups.

One could include the two-loop contribution in the matrix models which is expected to improve the high temperature behavior. If we use the same non-ideal form as given in (\ref{newnpt}) and assume the two-loop matrix model is obtained from (\ref{newmodel}) by including an overall factor $(1- 5 N g^2/(16\pi^2))$, an optimal fit shows that the values of the parameters are essentially unchanged as compared to table \ref{para} while the coupling $g^2$ is extremely small. The same can be found when using the two-parameter model in Ref.\cite{Dumitru:2012fw}. It turns out in our approach, the two-loop corrections are very small and negligible.

On the other hand, for a realistic value of $g^2$, one has to refit these parameters in the model and our results show a faster increase of the rescaled trace anomaly above $8T_c$ which indicates a better agreement with the lattice data as compared to the one-loop model. However, the behavior in the semi-QGP region is not satisfactory. It suggests that in order to reproduce the thermodynamics in both semi-QGP and perturbative region, the proper form of the non-ideal contributions that can accommodate the two-loop corrections should be considered. Furthermore, the two-loop corrections do not change the distribution of the eigenvalues of the Polyakov loop, one has to add the non-ideal terms to generate a phase transition. Therefore, the calculation at three-loop order is also important which is expected to answer if or not the distribution of eigenvalues can be modified. We postpone these studies to our future work.

\section*{Acknowledgements}
I would like to thank Adrian Dumitru and Rob Pisarski for their comments on the manuscript.
This work is supported by the NSFC of China under Project No.~11205035, by Natural Science Foundation of Guangxi Province of China under Project No.~2013GXNSFCA019002 and by Guangxi Normal University under Project No.~2011ZD004. Part of work is also supported by the European Research Council Grant No. HotLHC ERC-2001-StG-279579.

\appendix

\section{Bernoulli Polynomials}
\la{appa}

We define the Bernoulli polynomials,
\ba
\widehat B_{d-2k}(x)&=&T\sum_{n_0}\int {d^{d-1}\vec
  p\over{(2\pi)^{d-1}}} {1\over{(p^{ij})^{2k}}}\,,\\
\widehat B_{d-2k+1}(x)&=&T\sum_{n_0}\int {d^{d-1}\vec
  p\over{(2\pi)^{d-1}}} {p_0^{ij}\over{(p^{ij})^{2k}}}\,,\\
\widehat B_d(x)&=&T\sum_{n_0}\int {d^{d-1}\vec p\over{(2\pi)^{d-1}}}
\bigg(\log (p^{ij})^2 - \log p^2\bigg)\, .
\la{bernoulli}
\ea
In these equations, $p_0^{ij}=2\pi T (n_0 + x^{ij})$ with $n_0$ an
integer. Below, the indices $i$ and $j$ associated with $x$ are
omitted for simplicity of notation. Also, $(p^{ij})^2=(p_0^{ij})^2+{\vec
  p}^{\,2}$ and $p^2=(2\pi n_0 T)^2+{\vec p}^{\,2}$.

In $d=4$ dimensions and for $k=0$, 1, we have the following four
Bernoulli polynomials:
\ba
\widehat B_4(x)&=&{2\over 3}\pi^2T^4 B_4(x)\, ,\nonumber\\
\widehat B_3(x)&=&{2\over 3}\pi T^3 B_3(x)\, ,\nonumber\\
\widehat B_2(x)&=&{1\over 2}T^2 B_2(x)\, ,\nonumber\\
\widehat B_1(x)&=&-{T\over{4\pi}} B_1(x)\, ,
\la{bernoulli2}
\ea
with
\ba
B_4(x)&=&x^2(1-x)^2\, ,\nonumber\\
B_3(x)&=&x^3-{3\over 2}x^2+{1\over 2}x\, ,\nonumber\\
B_2(x)&=&x^2-x+{1\over 6}\, ,\nonumber\\
B_1(x)&=&x-{1\over 2}\, .
\la{bernoulli3}
\ea
The above expressions are defined on the interval $0\le x \le1$ and
they are periodic functions of $x$, with period $1$. For arbitrary
values of $x$, the argument of the above Bernoulli polynomials should
be understood as $x-[x]$ with $[x]$ the largest integer less than or
equal to $x$, which is nothing but the modulo function.

If $-1\le x \le1$ we can drop the modulo functions
and the Bernoulli polynomials reduce to
\ba
B_4(x)&=&x^2(1-\epsilon(x)x)^2\, ,\nonumber\\
B_3(x)&=&x^3-{3\over 2}\epsilon(x)x^2+{1\over 2}x\, ,\nonumber\\
B_2(x)&=&x^2-\epsilon(x)x+{1\over 6}\, ,\nonumber\\
B_1(x)&=&x-{1\over 2}\epsilon(x)\, ,
\la{bernoulli4}
\ea
where $\epsilon(x)$ is the sign function.

%Eq.~(\ref{bernoulli4}) is used to compute the effective potential.
%In fact the Bernoulli polynomials $B_1(x)$ and $B_3(x)$ are odd functions of $x$, while $B_2(x)$ and $B_4(x)$ are even
%functions of $x$, so we can always make the arguments of Bernoulli polynomials
%positive(or be zero) and ignore the sign functions which can save a lot of
%computing time. However, we point out that $B_1(x)$ has discontinuities at integer $x$. For
%example, the value of $B_1(0)$ depends on the way one approaches zero,
%from above or from below. If the result of the effective potential depends on $B_1(0)$, we have to know
%how one approaches zero in order to use the correct values of $B_1(0)$. In this case, the sign function in
%$B_1(x)$ is very important and can not be dropped even $x\ge 0$.
%Fortunately, we can prove that the contributions
%related to $B_1(n_0)$ vanish without specifying the value of
%$B_1(n_0)$. Therefore, the effective potential does not depend on
%$B_1(n_0)$ and we can simply drop the sign functions when $x\ge 0$. The proof is straightforward when using the
%total antisymmetry of the structure constants. Alternatively, one can
%also prove it by using the simplified expressions of the insertion
%diagram given in Sec.~\ref{sim}.

\section{Proof of Equation (\ref{id3})}
\la{proofid3}

In this appendix, we prove the identity given in (\ref{id3}). To make the proof more clear, we can use the periodicity of the Bernoulli polynomials to impose a constraint on the values of the background field $q_i$. Without losing any generality, we can assume $0 \le q_i <1$ thanks to the following properties of the modulo operation
\ba
&&{\rm Mod}[(q_i+n_i)-(q_j+n_j), 1] = {\rm Mod}[(q_i-q_j)+(n_i-n_j),1]\nonumber \\
&&\quad={\rm Mod}[{\rm Mod}[q_i-q_j,1]+{\rm Mod}[n_i-n_j,1],1]={\rm Mod}[q_i-q_j,1]\, .
\ea
Here, $n_i$ is an integer which satisfies $0 \le q_i+n_i <1$. $n_j$ is defined similarly. With our constraint of the values of the background field, the argument of the Bernoulli polynomials is in the interval $(-1, 1)$.

In principle, we can directly use (\ref{bernoulli4}) to prove (\ref{id3}). However, there is a problem related to $\widehat B_1(x)$ because it has discontinuities at integer $x$. The value of $\widehat B_1(x)$ at $x=0$ depends on the way how $x$ approaches zero, from above
or from below\footnote{With our constraint on the background field, the argument $x$ of the Bernoulli polynomials satisfies $-1<x<1$. So we only need to consider the discontinuity at $x=0$ for $\widehat B_1(x)$.}. We have $\widehat B_1(0^+)=-\frac{1}{2}$ and $\widehat B_1(0^-)=\frac{1}{2}$. Due to the discontinuity,  $q_i=q_j$ should be understood as $q_i=q_j\pm \epsilon$. Here, $\epsilon$ is an infinitely small (positive) number and the $+$ sign corresponds to $q_i$ approaches $q_j$ from above, while $-$ sign is for approach from below. This problem of discontinuity seems to make the discussion very complicated if some of the background fields are equal. Fortunately, we find that, in order to prove (\ref{id3}), there is no need to specify a way how $q_i$ approaches $q_j$, although the value of ${\cal F}(q_i,q_j,q_k)$ depends on this specification\footnote{Of course, (\ref{id3}) does not depend on this specification as we will see later.}.

On the other hand, with the condition $q_i\neq q_j\neq q_k$, it is obvious that we will not encounter the problem of discontinuity as we discussed above. we can directly use (\ref{bernoulli4}) to prove (\ref{id3}) by assuming a specific ordering of these background fields, for example, $q_i < q_j < q_k$.\footnote{The only reason to use a specific ordering of these background fields is to make the sign functions explicit, then the proof becomes straightforward. There are six possible ways to order the values of $q_i$, $q_j$ and $q_k$. Of course, the result does not depend on the orderings.}

Let's consider the second case where $q_i=q_j=q_k$. Although $\widehat B_1(q_i-q_j)$ is not determined until one specifies a way how $q_i$ approaches $q_j$, the product $\widehat B_1(q_i-q_j)\widehat B_3(q_i-q_k)$  is zero independent on the specification. This is because $\widehat B_3(0)=0$. Actually, it is easy to check that both the left and the right sides of (\ref{id3}) are $\frac{T^4}{48}$ in this case. For the same reason, we don't need a specification on how $q_i$ approaches $q_j$ in (\ref{id2}).

The last case corresponds to only two of the three background fields are equal. For example, we consider $q_i=q_j$. The following discussion simply applies for $q_i=q_k$ and $q_k=q_j$.

Explicitly, we have
\ba
{\cal F}(q_i,q_j,q_k) &=& 2 \widehat B_1(q_i-q_j)\widehat B_3(q_i-q_k)+ \widehat B_2(0)\widehat B_2(q_i-q_k)\,,\nonumber \\
{\cal F}(q_j,q_k,q_i) &=& 2 \widehat B_1(q_j-q_i)\widehat B_3(q_i-q_k)+ \widehat B_2(0)\widehat B_2(q_i-q_k)\,,\nonumber \\
{\cal F}(q_k,q_i,q_j) &=& 4 \widehat B_1(q_k-q_i)\widehat B_3(q_k-q_i)+ (\widehat B_2(q_i-q_k))^2\,.
\ea
Using the identity
\be
\la{b2can}
(\widehat B_1(0^+)+\widehat B_1(0^-))\widehat B_3(q_i-q_k) =0\,,
\ee
we find that the first term in ${\cal F}(q_i,q_j,q_k)$ and ${\cal F}(q_j,q_k,q_i)$ are cancelled by each other. Therefore, the left side of (\ref{id3}) becomes
\ba
{\cal F}(q_i,q_j,q_k)+{\cal F}(q_j,q_k,q_i)+{\cal F}(q_k,q_i,q_j)&=&4 \widehat B_1(q_k-q_i)\widehat B_3(q_k-q_i)+ 2 \widehat B_2(0)\nonumber \\&&\widehat B_2(q_i-q_k)+(\widehat B_2(q_i-q_k))^2\,.
\ea
Furthermore, the above result can be rewritten in terms of $\widehat B_4(q_k-q_i)$ with the help of (\ref{id1}) and (\ref{id2}),
\be
{\cal F}(q_i,q_j,q_k)+{\cal F}(q_j,q_k,q_i)+{\cal F}(q_k,q_i,q_j)= \frac{T^4}{48}-\frac{5}{8 \pi^2} \widehat B_4(q_k-q_i)\,.
\ee

Summing up the above discussions, we have proved (\ref{id3}). This identity is true for any value of the background fields and independent on the way how $q_i$ approaches $q_j$ in case these two fields are equal.

\end{document}